\documentclass[aps,superscriptaddress,nofootinbib,a4paper,longbibliography,twocolumn]{revtex4-2}
\pdfoutput=1
\usepackage[utf8]{inputenc}
\usepackage[T1]{fontenc}
\usepackage{amsmath,amssymb,mathtools,amsthm,bm,booktabs,graphicx,placeins,microtype}
\usepackage[colorlinks=true,allcolors=blue]{hyperref}
\providecommand{\doi}[1]{\href{https://doi.org/#1}{doi:#1}}

\newcommand{\Q}{\mathcal{Q}}
\newcommand{\Pguess}{P_{\mathrm{guess}}}
\newcounter{textbox}
\newtheorem{theorem}{Theorem}
\newtheorem{lemma}{Lemma}
\newtheorem{proposition}{Proposition}
\newtheorem{corollary}{Corollary}
\counterwithout{subsubsection}{subsection}
\counterwithin{subsubsection}{section}

\begin{document}

\title{Explicit construction of the normal cone at every nonlocal extreme behaviour of $\mathcal{Q}_{2222}$ and device-independent applications}

\author{Yutian Wu}
\author{Xie Sicheng}
\author{Ravishankar Ramanathan}
\email{ravi@cs.hku.hk}
\affiliation{School of Computing and Data Science,
The University of Hong Kong, Pokfulam Road, Hong Kong}

\begin{abstract}
The set of quantum correlations $\mathcal{Q}_{2222}$, in the simplest Bell scenario with two parties performing two binary measurements each, is an object of fundamental interest both in quantum foundations and for device-independent applications. A nonlocal extremal quantum behaviour in this set can maximize many different Bell expressions, yet these expressions are not all equally useful in a practical experiment. Here, firstly, we give an explicit characterization of every Bell expression maximized by any given
nonlocal extremal quantum behaviour in this Bell scenario.
This determines the full normal cone at every
such behaviour and makes it possible to optimize
over all Bell tests with the same ideal behavior.
We then derive an exact variational formula for
the quantum bound of an arbitrary Bell expression in this scenario.
As an example illustrating the construction, we derive the normal cone for a family of extremal quantum behaviours of importance in device-independent randomness
certification~\cite{WBC2022}. Finally, we explore the consequences of this characterization for device-independent randomness generation in practical experiments. Specifically, for any given detection efficiency $\eta \in [0.75,0.98]$, we construct Bell tests and
two-qubit experiments that certify more single-round min-entropy from one Bell score
than any CHSH score can certify with the same
detector efficiency, even when the quantum
devices used for CHSH are optimized.
The results provide a systematic way to choose
Bell tests for a desired quantum behaviour
and to improve their performance for device-independent applications under experimental imperfections.
\end{abstract}

\maketitle

\section{Introduction}
\label{sec:intro}

Quantum nonlocality reveals a fundamental distinction
between quantum theory and local hidden-variable
models: measurements on separated quantum systems can produce
correlations that violate Bell
inequalities~\cite{Bell1964,CHSH1969,Brunner2014}.
Beyond its foundational significance, the phenomenon of quantum nonlocality is the resource that enables device-independent information
processing. A Bell violation allows properties of an
experiment to be certified from its observed
correlations, without assuming a detailed model of the
state or measurement devices. This principle underlies
device-independent
randomness generation~\cite{Pironio2010},
quantum key distribution~\cite{ArnonFriedman2018},
and randomness amplification~\cite{ColbeckRenner2012,
Brandao2016}, and enables the certification of
entangled states and measurements through
self-testing~\cite{Brunner2014}.
Understanding which correlations quantum theory
permits, and which Bell expressions certify their
properties, is therefore of high significance both for the foundations of quantum theory and for its applications.

The simplest setting in which Bell nonlocality occurs
has two parties, each choosing between two measurements
with two outcomes. Its quantum set,
denoted $\mathcal Q_{2222}$, occupies a central position in
both research directions. Foundationally, it provides the
minimal setting in which to study the boundary between
classical, quantum, and general no-signalling
correlations. It includes the Tsirelson bound for the well-known CHSH inequality, while already exhibiting a nontrivial
geometry of extreme points and supporting Bell
inequalities~\cite{CHSH1969,Tsirelson,GohGeometry,TPLe2023, ZRLH23, Ramanathan2018}.
Operationally, the same small number of inputs and
outputs makes this scenario the most well-studied and practical one for
randomness certification and self-testing, including
protocols designed for inefficient
detectors~\cite{Pironio2010,WBC2022,Gigena2025}.
Furthermore, an important reduction of the nonlocal extreme behaviours in this set to
two-qubit realizations makes detailed analysis
possible~\cite{Jordan,Masanes}.

The characterization of $\mathcal Q_{2222}$ has
accordingly attracted sustained attention, from
quantum bounds and correlation boundary constraints ~\cite{Tsirelson,Landau}, to
studies of extremality and the geometry of the full
set including marginal
expectations~\cite{Ishizaka,GohGeometry,
MikosKaniewski,Ishizaka2025}.
Recently, a significant advance has been made in the form of the characterization of the extremal points by Barizien and
Bancal~\cite{BBQuantumStats}, the exact identification
of quantum correlations by Sun, Wang, and
Yu~\cite{SunWangYu2026}, and the description by one
of us in terms of no-disturbance-without-uncertainty
(NDWU) constraints and classical
randomization~\cite{RamanathanNDWU, SunZhouYu}.
In~\cite{BBExtremal}, Barizien and
Bancal studied the dual question regarding which Bell expressions are maximized by a given extremal behavior, and determined the full octagonal supporting
family at the Tsirelson point~\cite{BBExtremal}.
Sun, Wang, and Yu also proved that every Bell functional with a strict quantum advantage has a unique maximizing quantum behaviour, and analysed normal cones at nonlocal exposed points, including a complete cone certified in an explicit example~\cite{SunWangYu2026}. Recent work by Liu et al.\ established uniqueness and self-testing for Bell inequalities with a quantum advantage, proving that the normalized supporting families at any nonlocal
extreme behaviour have dimension at most two, and
explicitly determined these families along a
partially entangled curve~\cite{LiuZingarofaloBB2026}.

\begin{figure*}[t]
\centering
\includegraphics[width=0.62\textwidth]{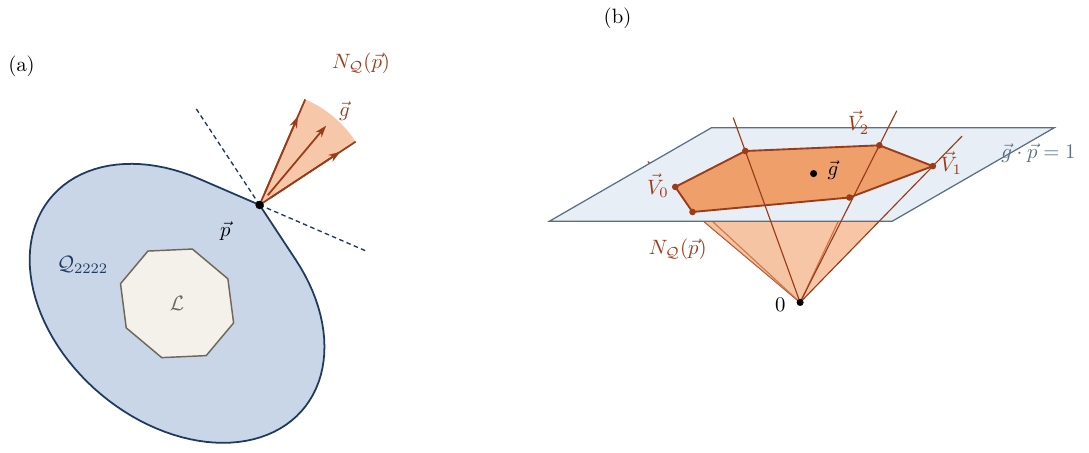}
\caption{The normal cone at a nonlocal extreme
behaviour of $\mathcal{Q}_{2222}$.
(a)~In a two-dimensional section of the space of
behaviours, $\vec p$ is an extreme point of the
quantum set, lying outside the local set $\mathcal{L}$.
Every Bell expression that attains its quantum
maximum at $\vec p$ is an outward normal there,
and these normals form the cone $N_{\mathcal{Q}}(\vec p)$.
(b)~In the space of Bell expressions, that cone
is generated from the origin. Its intersection
with the plane $\vec g\cdot\vec p=1$ is a polygon, each point of the
polygon is a supporting expression.}
\label{fig:normalcone}
\end{figure*}

These developments motivate the characterisation of the normal cone for arbitrary extremal quantum behaviors in $\mathcal{Q}_{2222}$. That is, given an arbitrary extremal quantum point (an ideal behavior that an experimentalist might aim to implement), which Bell
expressions attain their quantum maximum at that
behaviour? The complete answer is its
normal cone, the family of all supporting
linear functionals, illustrated in
Fig.~\ref{fig:normalcone}. Knowing this cone describes the freedom in a Bell test after its extremal
behaviour has been fixed. That freedom can have
operational consequences. Different supporting
expressions can respond differently to detector
loss, noise or other imperfections, and different scores
evaluated on the same recorded behaviour can certify
different amounts of
randomness~\cite{NietoSilleras2014,Bancal2014}.
A complete description thus allows optimization over all
target-supporting expressions, and can certify that no other expression in that family performs better for a
specified objective. The central difficulty is
global, i.e., conditions derived at the target realization
must exclude a larger Bell value at every other
quantum realization.

In this work, we give an explicit
necessary-and-sufficient construction of the full
normal cone at every nonlocal extreme behaviour of
$\mathcal Q_{2222}$. To do this, we build on the characterisation from \cite{RamanathanNDWU, SunZhouYu, BBExtremal}. Four tangent expressions,
associated with the states prepared on Bob's side
by Alice's inputs and outcomes, generate the
fixed-overlap supporting cone. We prove that a
nonnegative combination supports the whole quantum
set precisely when it satisfies one
measurement-stationarity equation and comparison
with the sixteen local deterministic behaviours.
Our proof uses the measurement-overlap determinant
and its correlator-product leading coefficient,
established in Proposition 4 of Ref.~\cite{SunWangYu2026} and also used in the uniqueness argument
of Ref.~\cite{LiuZingarofaloBB2026}, to establish
the general sufficiency statement.
We then use the same conditional-state description to obtain an
exact variational formula for quantum values for arbitrary Bell expressions in this scenario. We illustrate the construction by giving the normal cone for a specific family of points that are important for  device-independent (DI) randomness generation found by Wooltorton,
Brown, and Colbeck in~\cite{WBC2022}, obtaining a quadrilateral normalized section of the normal cone at every member of the family.
Finally, we identify an application of the construction of the normal cone for DI randomness generation. Specifically, for any given practical detection efficiency $\eta \in [0.75,0.98]$, we construct Bell tests and two-qubit experiments that certify more single-round min-entropy from one Bell score
than any CHSH score can certify with the same
detector efficiency, even when the quantum
devices used for CHSH are optimized. In this application, the detector model specifies only
the honest experiment; the randomness certificates
hold against general quantum side information (arbitrary quantum devices compatible
with an observed Bell value). 


The paper is organized as follows.
Section~\ref{sec:cone} introduces the normal-cone
construction and the exact quantum-bound formula.
Section~\ref{sec:idelta} determines the supporting
quadrilaterals of the $I_\delta$ family of~\cite{WBC2022}.
Section~\ref{sec:lossy} develops the detector-loss
application and its comparison with the CHSH
benchmark, and Section~\ref{sec:discussion}
discusses the implications, open questions and future directions.

\section{The normal cone at extremal quantum behaviors}
\label{sec:cone}
\label{sec:notation}
\label{sec:dual}

\subsubsection{Background}
We consider a Bell experiment in which Alice and Bob each choose one of two inputs,
$x,y\in\{0,1\}$, and obtain binary outcomes
$a,b\in\{-1,+1\}$. A behaviour is the collection of
conditional probabilities $P(ab|xy)$ of these
outcomes. The behaviour is no-signalling when Alice's outcome probabilities do not depend on Bob's choice of input,
and conversely. Explicitly,
$\sum_b P(ab|xy)$ is independent of $y$, and
$\sum_a P(ab|xy)$ is independent of $x$.
We write $\mathrm{NS}$ for the set of all
no-signalling behaviours in this Bell scenario.

A behaviour is said to be local realistic/classical when it can be realized by
assigning, in advance, a definite outcome to each of
Alice's two inputs and each of Bob's two inputs, or
by randomly choosing among such assignments.
Each of the four inputs is assigned an outcome
$\pm1$, so there are $2^4=16$ local deterministic
assignments. 
We write $\mathrm L$
for this set.

A behaviour is quantum when it can be realized by a
shared quantum state and a two-outcome measurement
for each input on each side. Explicitly, there exist
Hilbert spaces for Alice and Bob, a state $\rho$ on
their joint space, and observables $A_x$
($x=0,1$) and $B_y$ ($y=0,1$), each with eigenvalues
$\pm1$, such that
\begin{equation}
\label{eq:Born}
P(ab|xy)
=\mathrm{Tr}\Bigl[
\rho\,
\Bigl(\frac{I+aA_x}{2}\Bigr)
\otimes
\Bigl(\frac{I+bB_y}{2}\Bigr)
\Bigr].
\end{equation}
Here $A_x$ is the observable measured by Alice when
her input is $x$, and $B_y$ is the observable
measured by Bob when his input is $y$. We write
$\mathcal Q=\mathcal Q_{2222}$ for the set of all
behaviours of this form. These sets satisfy
$\mathrm L\subset\mathcal Q\subset\mathrm{NS}$ \cite{Brunner2014}.

A behaviour is nonlocal if it does not belong to
$\mathrm L$. A quantum behaviour is said to be extremal if it
cannot be obtained by randomly mixing two different
quantum behaviours, i.e., whenever
$\vec p=\lambda\vec p'+(1-\lambda)\vec p''$
with $0<\lambda<1$ and
$\vec p',\vec p''\in\mathcal Q$, one has
$\vec p'=\vec p''=\vec p$.
This paper concerns the nonlocal extreme
points of $\mathcal Q$.

Any no-signalling behaviour is determined by eight
real numbers, which are collected in the vector
\begin{equation}
\label{eq:pvec}
\vec p
:=(\alpha_0,\alpha_1,\beta_0,\beta_1,
E_{00},E_{01},E_{10},E_{11}).
\end{equation}
Here
$\alpha_x=\sum_{a,b}a\,P(ab|xy)$
is Alice's outcome bias for input $x$, independent
of $y$ by no-signalling;
$\beta_y=\sum_{a,b}b\,P(ab|xy)$
is Bob's outcome bias for input $y$;
and
$E_{xy}=\sum_{a,b}ab\,P(ab|xy)$
is the correlator for the input pair $(x,y)$.
Equivalently,
\begin{equation}
\label{eq:Pab}
P(ab|xy)=\frac14
\left(1+a\alpha_x+b\beta_y+abE_{xy}\right).
\end{equation}
In a quantum realization these components are the
expectation values
$\alpha_x=\mathrm{Tr}\bigl[\rho\,(A_x\otimes I)\bigr]$,
$\beta_y=\mathrm{Tr}\bigl[\rho\,(I\otimes B_y)\bigr]$,
and
$E_{xy}=\mathrm{Tr}\bigl[\rho\,(A_x\otimes B_y)\bigr]$.

A Bell expression is then a linear functional of this
vector, i.e.,
\begin{equation}
\label{eq:Bell}
\mathcal B_{\vec g}(\vec p)
=\vec g\cdot\vec p
=\sum_x g_{\alpha_x}\alpha_x
+\sum_y g_{\beta_y}\beta_y
+\sum_{x,y}g_{E_{xy}}E_{xy}.
\end{equation}
The eight numbers
$g_{\alpha_x}$, $g_{\beta_y}$, and $g_{E_{xy}}$ for $x, y \in \{0,1\}$ constitute the coefficient vector of the Bell expression 
\begin{equation}
\label{eq:gvec}
\vec g
:=(g_{\alpha_0}, g_{\alpha_1}, g_{\beta_0}, g_{\beta_1},
g_{E_{00}}, g_{E_{01}}, g_{E_{10}}, g_{E_{11}}).
\end{equation}
The local and quantum bounds of the Bell expression (denoted by its coefficient vector $\vec g$) are denoted as
\begin{equation}
\label{eq:hLhQ}
h_{\mathrm L}(\vec g):=\max_{\vec p\in\mathrm L}\vec g\cdot\vec p,
\qquad
h_{\mathcal Q}(\vec g):=\max_{\vec p\in\mathcal Q}\vec g\cdot\vec p.
\end{equation}
The local bound is attained at one of the sixteen
local deterministic assignments, since a linear function
on a polytope attains its maximum at a vertex.
The quantum bound is the maximum over all shared
states and all choices of the observables $A_x$ and
$B_y$.

We say that a Bell expression $\vec g$ supports the
quantum set at a behaviour $\vec p$ when
$\vec g\cdot\vec p=h_{\mathcal Q}(\vec g)$, that is,
when $\vec p$ attains the quantum bound of $\vec g$.
All such expressions form the normal cone
\begin{equation}
\label{eq:NQ}
N_{\mathcal Q}(\vec p)
:=\{\vec g \; : \; \vec g\cdot\vec p=h_{\mathcal Q}(\vec g)\}.
\end{equation}
Note that positive rescaling preserves this property, as does adding a constant to the Bell expression. We omit additive constants and remove the rescaling
freedom by fixing the target value to one. 
We also note that a supporting expression need not have a unique
maximizer or a strict quantum advantage.

\subsubsection{The normal cone for every nonlocal extremal point of $\mathcal{Q}_{2222}$}

Fix a nonlocal extremal behaviour $\vec p \in \mathcal{Q}_{2222}$. When Alice
measures her observable $A_x$ and obtains outcome $a$, she prepares
a conditional state on Bob's side. There are four
such states, one for each pair $(a,x)$. The
probability of this event is
$(1+a\alpha_x)/2$. If this probability vanished,
Alice would have a deterministic outcome for input
$x$. In that case a local model would exist in this minimal Bell scenario, explicitly one can
sample the outcome of her remaining measurement and
use Bob's response probabilities conditional on her
outcomes. A nonlocal behaviour therefore has
$(1+a\alpha_x)/2>0$ for every $a$ and $x$, and the
four conditional expectations below are well defined.

Bob's expectation of $B_y$ in the state prepared
when Alice obtains $a$ on input $x$ is the
conditional average  given as \cite{SunZhouYu,RamanathanNDWU}
\begin{equation}
\label{eq:w}
w_{a|x,y}
=\frac{\sum_b b\,P(ab|xy)}{P(a|x)}
=\frac{\beta_y+aE_{xy}}{1+a\alpha_x}.
\end{equation}
The second equality is from Eq.~\eqref{eq:Pab}, the numerator is $(\beta_y+aE_{xy})/2$ and the
denominator is $(1+a\alpha_x)/2$. We denote by
$\vec w_{a|x}=(w_{a|x,0},w_{a|x,1})$ the pair of these expectations.

In this Bell scenario, a well-known result by Masanes states that for extremal behaviors it suffices to consider two-qubit
realizations~\cite{Jordan,Masanes}. In particular, Bob's system may be taken to be a qubit, and each of his observables $B_y$ may be written as
\begin{equation}
\label{eq:By}
B_y=\vec n_y\cdot\boldsymbol{\sigma}.
\end{equation}
Here $\boldsymbol{\sigma}=(\sigma_x,\sigma_y,\sigma_z)$ denotes the three Pauli matrices,
and $\vec n_y$ is a real unit vector in three-dimensional space that specifies the direction of Bob's measurement for input $y$. 
Let $c=\vec n_0\cdot\vec n_1$ denote the overlap of these two directions, that is, the cosine of the angle between $\vec n_0$ and $\vec n_1$.

After Alice performs her measurement, Bob's qubit is left in a state described by its Bloch vector $\vec m=(m_x,m_y,m_z)$.
The expectation value of $B_y=\vec n_y\cdot\boldsymbol{\sigma}$ in this state is $\vec m\cdot\vec n_y$.
The four conditional states prepared by Alice each have their own Bloch vector, and therefore their own pair of expectations $\vec w_{a|x} :=(w_{a|x,0},w_{a|x,1})$. As $\vec m$ ranges over all vectors of length at most one, the possible pairs $\bigl(w_{a|x,0} = \vec m\cdot\vec n_0, w_{a|x,1} = \vec m\cdot\vec n_1\bigr)$ constitute an ellipse (together with its interior). For $|c|<1$ this ellipse is nondegenerate.
Explicitly, let $G_c$ be the $2\times2$ matrix of inner products of the two directions $\vec n_0$ and $\vec n_1$, namely
\begin{equation}
\label{eq:Gc}
G_c=\begin{pmatrix}1&c\\c&1\end{pmatrix},
\end{equation}
and write $\vec w=(\vec m\cdot\vec n_0,\vec m\cdot\vec n_1)$ for a generic pair of these two expectations. Let $\vec m_\parallel$ be the orthogonal
projection of $\vec m$ onto the plane spanned
by $\vec n_0$ and $\vec n_1$. Then
\[
\vec w^{\mathsf T}G_c^{-1}\vec w
=\lvert\vec m_\parallel\rvert^2
\leq\lvert\vec m\rvert^2\leq1.
\]
Each of the four conditional pairs therefore obeys
\begin{equation}
\label{eq:ndwu}
\vec w_{a|x}^{\mathsf T}G_c^{-1}\vec w_{a|x}\leq1.
\end{equation}
This is the uncertainty constraint for two
measurements that do not commute:
the two
expectations cannot both attain the values they
could if the measurements were compatible.

A two-qubit realization of $\vec p$ therefore
determines an overlap $c=\vec n_0\cdot\vec n_1$. If $c=\pm1$, Bob's
measurements are parallel or antiparallel and
commute, and every behaviour realizable at this
overlap is local. Since $\vec p$ is nonlocal, it
already belongs to some overlap in $(-1,1)$. The
next lemma shows that this overlap is unique.

\begin{lemma}[Unique measurement overlap]
\label{lem:unique-overlap}
Let $\vec p$ be a nonlocal extremal behaviour in
$\mathcal{Q}_{2222}$. There is a unique overlap
$c_0\in(-1,1)$ at which $\vec p$ can be realized by
a two-qubit strategy in which Bob's two measurement
directions $\vec n_0, \vec n_1$ have overlap $c_0 = \vec n_0 \cdot \vec n_1$.
\end{lemma}

\begin{proof}
Existence of at least one such overlap in $(-1,1)$
was just noted. To rewrite the uncertainty constraint
\eqref{eq:ndwu} as a condition on the overlap, invert the matrix of
Eq.~\eqref{eq:Gc}: for $|c|<1$,
\[
G_c^{-1}
=\frac{1}{1-c^2}
\begin{pmatrix}1&-c\\-c&1\end{pmatrix}.
\]
For each of the four pairs
$\vec w_{a|x}=(w_{a|x,0},w_{a|x,1})$, the inequality
of Eq.~\eqref{eq:ndwu} is then equivalent to
$F_{a|x}(c)\le 0$, where
\begin{equation}
\label{eq:F}
F_{a|x}(c)
=
w_{a|x,0}^2
+w_{a|x,1}^2
+c^2
-2c\,w_{a|x,0}w_{a|x,1}
-1.
\end{equation}

Now suppose the same behaviour could also be realized
with a second overlap $d\in(-1,1)$, $d\neq c$. Then
$F_{a|x}(c)\le 0$ and $F_{a|x}(d)\le 0$ for every
$a$ and $x$. The algebraic identity
\[
F_{a|x}\Bigl(\frac{c+d}{2}\Bigr)
=
\frac{F_{a|x}(c)+F_{a|x}(d)}{2}
-\frac{(c-d)^2}{4}
\]
implies that $F_{a|x}$ is strictly negative at the
midpoint $r=(c+d)/2$. This holds for all four pairs,
so they all lie strictly inside the ellipse belonging
to the intermediate overlap $r$.

Because these four inequalities are strict, they
remain valid when the coordinates of $\vec p$ are
varied slightly. No nonlocal quantum behaviour is a
vertex of the no-signalling set \cite{RTHH16}, so one may make this
variation among nearby no-signalling behaviours that
keep every vanishing probability of $\vec p$ equal to
zero. All probabilities then remain nonnegative.
Every behaviour in a small neighbourhood of this kind
still satisfies the four fixed-overlap constraints at
overlap $r$, and is therefore quantum, by the NDWU
characterization of Ref.~\cite{RamanathanNDWU}, as
recalled in Appendix~\ref{app:dual}. In particular,
a short line segment through $\vec p$ consists of
quantum behaviours, so $\vec p$ is a mixture of two
different quantum behaviours, contradicting
extremality. Hence only one overlap is possible. We
denote it by $c_0$.
\end{proof}

\begin{lemma}[Saturation of the uncertainty constraints]
\label{lem:saturation}
At the unique overlap $c_0$ of
Lemma~\ref{lem:unique-overlap}, each of the four
conditional pairs lies on the boundary of the
ellipse:
\begin{equation}
\label{eq:sat}
\vec w_{a|x}^{\mathsf T}G_{c_0}^{-1}\vec w_{a|x}=1
\qquad\text{for every }a,x.
\end{equation}
\end{lemma}

\begin{proof}
The inequality $\le 1$ is the uncertainty constraint
of Eq.~\eqref{eq:ndwu} at $c_0$. It remains to rule
out a strict inequality, that is, a pair lying in the
interior of the ellipse. By the two-qubit reduction~\cite{Jordan,Masanes},
the behaviour admits a two-qubit realization with
projective measurements. Choose local bases in
which both observables on each side are real,
and hence lie in the $\sigma_x$--$\sigma_z$ plane.
Replacing the shared density matrix $\rho$ by
$(\rho+\rho^*)/2$ preserves every observed
expectation value. This is a real positive
semidefinite matrix and therefore admits a
spectral decomposition into pure real states.
Since $\vec p$ is extreme, every positive-weight
component in this decomposition realizes
$\vec p$ with the same measurements. We may
therefore choose a pure real two-qubit state. Nonlocality excludes trivial projective
measurements, so Alice's outcome projectors
have rank one. Each of Bob's conditional states
is consequently pure and real. Its Bloch vector
$\vec m_{a|x}$ has unit length and lies in
Bob's measurement plane. Hence
\begin{equation}
\vec w_{a|x}^{\mathsf T}G_{c_0}^{-1}\vec w_{a|x}
=
\lvert\vec m_{a|x}\rvert^2
=1. \nonumber
\end{equation}
\end{proof}

The four points $\vec w_{a|x}$ therefore lie on the
ellipse at the unique overlap $c_0$. This ellipse is
the curve
$\vec w^{\mathsf T}G_{c_0}^{-1}\vec w=1$
in the plane of pairs of expectations
$\vec w=(w_0,w_1)$. Here $w_0$ and $w_1$ are simply
the two coordinates of a generic point in this
plane, while each
$\vec w_{a|x}=(w_{a|x,0},w_{a|x,1})$ is one of the
four specific points attached to $\vec p$.

At the point $\vec w_{a|x}$ on the ellipse, the
outward normal is obtained by differentiating the
quadratic form that defines the curve. The gradient
of $\vec w^{\mathsf T}G_{c_0}^{-1}\vec w$ with
respect to $\vec w$ is $2G_{c_0}^{-1}\vec w$, so the
outward normal at $\vec w_{a|x}$ is the
two-component vector
\begin{equation}
\label{eq:v}
\vec v_{a|x}
:=G_{c_0}^{-1}\vec w_{a|x}
=(v_{a|x,0},\,v_{a|x,1}).
\end{equation}
Because $\vec w_{a|x}$ lies on the
ellipse,
\[
\vec v_{a|x}\cdot\vec w_{a|x}
=\vec w_{a|x}^{\mathsf T}G_{c_0}^{-1}\vec w_{a|x}
=1.
\]
The tangent line at $\vec w_{a|x}$ therefore says
that every pair of expectations $\vec w=(w_0,w_1)$
allowed at overlap $c_0$ satisfies
\begin{equation}
\label{eq:tangent}
\vec v_{a|x}\cdot\vec w
\le
1,
\end{equation}
with equality at $\vec w=\vec w_{a|x}$.

This tangent inequality is a constraint on Bob's two
conditional expectations. Substituting the
expressions
$w_{a|x,y}=(\beta_y+a E_{xy})/(1+a\alpha_x)$ from
Eq.~\eqref{eq:w} into Eq.~\eqref{eq:tangent} gives
\[
\sum_y v_{a|x,y}\,
\frac{\beta_y+a E_{xy}}{1+a\alpha_x}
\le 1,
\]
where $v_{a|x,y}$ are the two components $v_{a|x,0}$ and $v_{a|x,1}$ of
$\vec v_{a|x}=(v_{a|x,0},\,v_{a|x,1})$. The denominator $1+a\alpha_x$ is
strictly positive at a nonlocal behaviour, as
already noted. Multiplying through by this factor
and rearranging terms, one obtains a linear
inequality in the coordinates of a behaviour,
\begin{equation}
\label{eq:tangentbell}
-a\,\alpha_x
+\sum_y v_{a|x,y}\,\beta_y
+\sum_y a\,v_{a|x,y}\,E_{xy}
\le 1.
\end{equation}
The left-hand side is the Bell value of a specific Bell
expression $\vec\gamma_{a|x}$. Explicitly, the only nonzero coefficients of this expression are those appearing
in Eq.~\eqref{eq:tangentbell}, given as
\begin{equation}
\label{eq:gamma}
\begin{aligned}
(\vec\gamma_{a|x})_{\alpha_x}&=-a,\\
(\vec\gamma_{a|x})_{\beta_y}&=v_{a|x,y},\\
(\vec\gamma_{a|x})_{E_{xy}}&=a\,v_{a|x,y}.
\end{aligned}
\end{equation}
There are four such expressions, one for each
conditional state labelled by $(a,x)$, and therefore one for each of the
four points $\vec w_{a|x}$ on the ellipse.

Let $K_{c_0}$ denote the set of quantum behaviours
that can be realized with Bob's measurement overlap
held fixed at this same value $c_0$. For every
behaviour $\vec z\in K_{c_0}$, the tangent inequality
just derived gives
\[
\vec\gamma_{a|x}\cdot\vec z\le 1,
\]
while at the target itself one has
$\vec\gamma_{a|x}\cdot\vec p=1$. Thus each of the
four expressions is maximized by $\vec p$ among all
quantum behaviours with overlap $c_0$. They need not
yet be maximized by $\vec p$ among the whole quantum
set, because a realization with a different overlap
might produce a larger value. These four expressions generate all Bell expressions
with that fixed-overlap property. Every Bell
expression $\vec g$ that is maximized by $\vec p$
among behaviours in $K_{c_0}$ can be written as a
nonnegative combination of the four tangent
expressions,
\begin{equation}
\label{eq:mu}
\vec g=\sum_{a,x}\mu_{a|x}\vec\gamma_{a|x},
\qquad \mu_{a|x}\geq0.
\end{equation}
The four numbers $\mu_{a|x}$ are the weights in this
combination: the weight $\mu_{a|x}$ multiplies the
tangent expression associated with the conditional
state of outcome $a$ on input $x$. Every such
combination remains maximized by $\vec p$ as long as
Bob's overlap is held at $c_0$. Its value at the
target is
$\vec g\cdot\vec p=\sum_{a,x}\mu_{a|x}$.

For $\vec g$ to support the whole quantum set at
$\vec p$, two further conditions are needed.
To express the first of these, we write a two-qubit
realization of $\vec p$ from the data already
obtained. Explicitly, choose two unit vectors $\vec n_0$ and
$\vec n_1$ in a plane with overlap
$\vec n_0\cdot\vec n_1=c_0$; one may take, for
instance, $\vec n_0=(1,0,0)$ and
$\vec n_1=(c_0,\sqrt{1-c_0^2},0)$. The four
conditional Bloch vectors are then
$\vec m_{a|x}=v_{a|x,0}\vec n_0+v_{a|x,1}\vec n_1$.
Bob's reduced state is the mixture of either pair
of these states, with weights $(1+a\alpha_x)/2$.
A purification of that reduced state, together with
the two projective measurements on Alice's qubit
that prepare these ensembles, is a realization of
$\vec p$, as guaranteed by the reduction of
Refs.~\cite{Jordan,Masanes}. Keep this shared
state and Alice's measurements fixed, and rotate
Bob's two measurement directions in their common
plane so that their overlap becomes a variable $c$.
At $c=c_0$ one recovers the original behaviour
$\vec p$. Because $c_0$ lies strictly between $-1$
and $1$, this rotation can be made in either
direction, and it produces a smooth curve
$\vec p(c)$ of quantum behaviours with
$\vec p(c_0)=\vec p$. Write
\begin{equation}
\label{eq:pdotdef}
\dot{\vec p}
=\left.\frac{d}{dc}\vec p(c)\right|_{c=c_0}
\end{equation}
for the tangent to this curve at the target. If
$\vec g$ is maximized by $\vec p$ among all quantum
behaviours, then the function
$f(c)=\vec g\cdot\vec p(c)$ has a maximum at
$c=c_0$, so its derivative must vanish there:
\begin{equation}
\label{eq:stat}
\vec g\cdot\dot{\vec p}=0.
\end{equation}
This is the stationarity condition. A different
realization may produce a different curve through
$\vec p$; the same necessary condition would then
apply along that curve as well. For the construction
it is enough to impose stationarity along the
explicit path just described. Second, the value at the target must not be exceeded by any local behaviour. The local set is the convex
hull of the sixteen local deterministic assignments, one
for each choice of outcomes
$a_0,a_1,b_0,b_1\in\{-1,+1\}$. The corresponding
behaviours are
\begin{equation}
\label{eq:detvec}
\vec d
=(a_0,a_1,b_0,b_1,
a_0b_0,a_0b_1,a_1b_0,a_1b_1).
\end{equation}
No local behavior can exceed the value at $\vec p$ if and only if 
\begin{equation}
\label{eq:det}
\vec g\cdot\vec d\le\vec g\cdot\vec p
\end{equation}
holds for all sixteen local deterministic behaviours
$\vec d$.

The combinations \eqref{eq:mu} that obey
stationarity \eqref{eq:stat} and the sixteen inequalities \eqref{eq:det} form
a cone: if $\vec g$ satisfies these conditions, then
so does every positive multiple of $\vec g$. We call
this cone the candidate cone $\mathcal C(\vec p)$ of $\vec p$,
\begin{equation}
\label{eq:cand}
\begin{aligned}
\mathcal C(\vec p)=\bigl\{\,
&\vec g=\sum_{a,x}\mu_{a|x}\vec\gamma_{a|x}:\\
&\mu_{a|x}\geq0,\quad \vec g\cdot\dot{\vec p}=0,\\
&\vec g\cdot\vec d\leq\vec g\cdot\vec p
\text{ for all local deterministic }\vec d
\bigr\}.
\end{aligned}
\end{equation}
Our first main result is then the following.

\begin{theorem}[Normal-cone construction]
\label{thm:dualcone}
For every nonlocal extreme behaviour
$\vec p\in\mathcal Q_{2222}$,
\begin{equation}
\label{eq:NQC}
N_{\mathcal Q}(\vec p)=\mathcal C(\vec p).
\end{equation}
\end{theorem}

\textit{Proof sketch.-}
We establish
necessity, then sufficiency.

\textit{Step 1.-}
If a Bell expression $\vec g$ supports the whole
quantum set at $\vec p$, then in particular it
supports the smaller set $K_{c_0}$ at $\vec p$. It
is therefore a nonnegative combination of the four
tangent expressions, of the form
Eq.~\eqref{eq:mu}.

\textit{Step 2.-}
Along the curve $\vec p(c)$ of quantum behaviours
obtained by varying Bob's overlap as above, the
Bell value $f(c)=\vec g\cdot\vec p(c)$ has a
maximum at $c=c_0$, so the stationarity condition
Eq.~\eqref{eq:stat} holds. Since every local
behaviour is quantum, the sixteen inequalities
Eq.~\eqref{eq:det} hold as well. Hence every
supporting expression lies in the candidate cone,
$N_{\mathcal Q}(\vec p)\subseteq\mathcal C(\vec p)$.

\textit{Step 3.-}
For the converse, take a nonzero
$\vec g\in\mathcal C(\vec p)$ and normalize its
value at the target to one. In a two-qubit representation,
the measurement operators are parametrized, up to
local changes of basis, by Alice's and Bob's
measurement overlaps. Let $H$ be the corresponding
Bell operator. The determinant
$\Delta=\det(I_4-H)$ is a polynomial of degree at
most two in each overlap, and the target is an
interior stationary zero of this polynomial.
Its leading coefficient has the sign of
$\prod_{x,y}g_{E_{xy}}$.
When this product is nonpositive, restricting
$\Delta$ to a line through the target gives
$\Delta(\sigma)=\sigma^2q(\sigma)$ with $q$
concave. The local comparisons imply
nonnegativity at the endpoints, where at least
one party's measurements commute, and therefore
throughout the line segment.
Thus $\Delta\geq0$ for all overlaps.
Under the same local comparisons,
Lemma~\ref{lem:upgrade} upgrades determinant
nonnegativity to $I_4-H\succeq0$.
This excludes a Bell value greater than one
for every state. 

\textit{Step 4.-}
If the product of the four correlator coefficients
is strictly positive, local outcome relabellings
make all four coefficients positive.
In these relabelled coordinates, a shared-threshold
local model preserves the target marginals and
makes every correlator no smaller, with at least
one strictly larger. Its Bell value is therefore
strictly larger than the target value, contradicting
the local comparison. 
Hence
$\mathcal C(\vec p)\subseteq N_{\mathcal Q}(\vec p)$.
\qed

The full proof is given in
Appendix~\ref{app:dualcone}. In words, Theorem \ref{thm:dualcone} states that the Bell expressions maximized by a given extremal
$\vec p$ are exactly the
nonnegative combinations of the four tangent
expressions that are stationary under the overlap
variation and that do not exceed their target value
at any of the sixteen local deterministic behaviours. After normalization by $\vec g\cdot\vec p=1$,
the construction gives a polytope of supporting
expressions. Its dimension is at most two,
as established in
Ref.~\cite[Lemma~4]{LiuZingarofaloBB2026}.

We collect the characterisation of the normal cone at a given extremal quantum point in the following box.

\begin{center}
\refstepcounter{textbox}%
\label{box:recipe}%
\fbox{%
\begin{minipage}{\dimexpr\columnwidth-2\fboxsep-2\fboxrule\relax}
\small
\setlength{\topsep}{0.3em}
\setlength{\itemsep}{0.28em}
\setlength{\parsep}{0pt}
\setlength{\abovedisplayskip}{0.4em}
\setlength{\belowdisplayskip}{0.4em}
\setlength{\abovedisplayshortskip}{0.15em}
\setlength{\belowdisplayshortskip}{0.15em}
\textit{Box~\thetextbox.} \textit{Computing the normal cone.-}
Given a nonlocal extreme behaviour
$\vec p=(\alpha_0,\alpha_1,\beta_0,\beta_1,
E_{00},E_{01},E_{10},E_{11})$, the normal cone $N_{\mathcal Q}(\vec p)$ at this behaviour is obtained as follows.
\begin{enumerate}
\item
For the given $\vec p$, compute the four
conditional pairs
$\vec w_{a|x}=(w_{a|x,0},w_{a|x,1})$ using
Eq.~\eqref{eq:w}. 
\item
Find the unique overlap $c_0\in(-1,1)$ at which
these four pairs lie on the ellipse, i.e.,
solve $F_{a|x}(c)=0$ for the quadratic of
Eq.~\eqref{eq:F}. By Lemma~\ref{lem:unique-overlap}
there is a single root $c_0$ that works for all
four pairs, and at that root Eq.~\eqref{eq:sat}
holds.
\item
Compute the four normals
$\vec v_{a|x}=G_{c_0}^{-1}\vec w_{a|x}$, and write
their components as
$\vec v_{a|x}=(v_{a|x,0},v_{a|x,1})$. Insert these
components into Eq.~\eqref{eq:gamma} to obtain the
four tangent expressions $\vec\gamma_{a|x}$.
\item
Using $c_0$ and the four normals
$\vec v_{a|x}$, write a two-qubit
realization of $\vec p$ as follows. Choose two unit
vectors $\vec n_0$ and $\vec n_1$ in a plane with
the given overlap $\vec n_0\cdot\vec n_1=c_0$. The four
conditional Bloch vectors are then
$\vec m_{a|x}=v_{a|x,0}\vec n_0+v_{a|x,1}\vec n_1$.
Bob's reduced state is the mixture of either pair
of these states, with weights $(1+a\alpha_x)/2$.
Each corresponding state on Bob's qubit is
$\rho_{a|x}=(I+\vec m_{a|x}\cdot\boldsymbol{\sigma})/2$,
and this reduced state may be written
\[
\rho_B=\sum_{a=\pm1}\frac{1+a\alpha_x}{2}\,\rho_{a|x},
\]
independently of $x$. Let
$\vec m_B=\sum_{a=\pm1}((1+a\alpha_x)/2)\,\vec m_{a|x}$
be its Bloch vector, and write $r=|\vec m_B|$.
If $\vec m_B\neq0$, let $|e_+\rangle$ and $|e_-\rangle$
be the qubit states with Bloch vectors $\vec m_B/r$
and $-\vec m_B/r$; if $\vec m_B=0$, take any
orthonormal pair in the plane, for instance the
eigenstates of $\vec n_0\cdot\boldsymbol{\sigma}$.
A shared state with this reduced state is
\[
|\psi\rangle
=\sqrt{\frac{1+r}{2}}\,|e_+\rangle_A|e_+\rangle_B
+\sqrt{\frac{1-r}{2}}\,|e_-\rangle_A|e_-\rangle_B.
\]
In the Schmidt basis used above, Alice's
measurement effects are
\[
\Pi_{a|x}
=\frac{1+a\alpha_x}{2}
\left(
\rho_B^{-1/2}\rho_{a|x}\rho_B^{-1/2}
\right)^{\mathsf T}.
\]
Here the transpose is taken in that basis,
with the resulting matrix acting on Alice's
qubit. Nonlocality implies that $\rho_B$ has
full rank, so the inverse is well defined.
For each input, these are two rank-one positive
operators summing to the identity, and hence
are orthogonal projectors.
The corresponding observable is
$A_x=\Pi_{+|x}-\Pi_{-|x}$.
This is a realization of $\vec p$. 
\item 
Keeping this state and Alice's
measurements fixed, rotate Bob's directions in the
plane spanned by $\vec n_0$ and $\vec n_1$ so as to vary the overlap, and differentiate
the resulting curve of behaviours at $c=c_0$. This
produces the vector $\dot{\vec p}$ of
Eq.~\eqref{eq:pdotdef}. Also list the sixteen local
deterministic behaviours $\vec d$ of
Eq.~\eqref{eq:detvec}.
\item
A general candidate Bell expression is a combination
$\vec g=\sum_{a,x}\mu_{a|x}\vec\gamma_{a|x}$ with
four weights $\mu_{a|x}\ge 0$. The normal cone
consists of those combinations that satisfy the
stationarity condition $\vec g\cdot\dot{\vec p}=0$
and the sixteen local inequalities
$\vec g\cdot\vec d\le\vec g\cdot\vec p$.
\end{enumerate}
\end{minipage}}
\end{center}

\subsubsection{An exact expression for the Quantum value of Bell expressions in the minimal scenario}

The same description of conditional states also
determines the exact quantum value of an arbitrary
Bell expression, including expressions that are not
already known to be supporting at a given point.
Let
\begin{equation}
\label{eq:gvecThm}
\vec g
=(g_{\alpha_0}, g_{\alpha_1}, g_{\beta_0}, g_{\beta_1},
g_{E_{00}}, g_{E_{01}}, g_{E_{10}}, g_{E_{11}})
\end{equation}
be a given Bell expression. Write
$\vec g_\beta=(g_{\beta_0},g_{\beta_1})$
and
$\vec g_{E_x}=(g_{E_{x0}},g_{E_{x1}})$
for the coefficients that multiply
Bob's two marginals and the two correlators on
Alice's input $x$ respectively. For any overlap $c\in[-1,1]$
and any two real numbers $\lambda_0$ and
$\lambda_1$, write
\begin{equation}
\label{eq:normc}
\|(\lambda_0,\lambda_1)\|_c
=\sqrt{\lambda_0^2+\lambda_1^2+2c\,\lambda_0\lambda_1}.
\end{equation}
The exact quantum bound of $\vec g$ is then given
by \eqref{eq:hQ} in the following theorem, in which the minimum
runs over all pairs
$\vec\delta=(\delta_0,\delta_1)$ of real numbers.

\begin{theorem}[Exact quantum value]
\label{thm:dualmain}
\label{thm:dual}
For every Bell expression $\vec g =(g_{\alpha_0}, g_{\alpha_1}, g_{\beta_0}, g_{\beta_1},
g_{E_{00}}, g_{E_{01}}, g_{E_{10}}, g_{E_{11}})$ in the Bell scenario (2,2;2,2), the quantum value is given as
\begin{equation}
\label{eq:hQ}
\begin{split}
h_{\mathcal Q}(\vec g)
={}&\max_{c\in[-1,1]}
\min_{\vec\delta\in\mathbb R^2}
\sum_{x=0}^{1}\max_{a=\pm1}\\
&\left[
a g_{\alpha_x}+
\left\|\frac{\vec g_\beta}2+(-1)^x\vec\delta+a\vec g_{E_x}\right\|_c
\right].
\end{split}
\end{equation}
\end{theorem}

\textit{Proof sketch.-}

\textit{Step 1.-}
Every quantum behaviour is a convex combination of
behaviours realized at a single fixed overlap of
Bob's measurements. The quantum bound of $\vec g$
is therefore the largest of the bounds obtained when
the overlap is held fixed at a value $c\in[-1,1]$.

\textit{Step 2.-}
At a fixed $c$, the Bell value of
$\vec g$ is at most
\[
\sum_{x=0}^{1}\max_{a=\pm1}
\left[
a g_{\alpha_x}+
\left\|\frac{\vec g_\beta}2+(-1)^x\vec\delta+a\vec g_{E_x}\right\|_c
\right],
\]
for every pair $\vec\delta=(\delta_0,\delta_1)$.
Each of the four terms corresponds to one pair
$(a,x)$ of Alice's input and outcome. The length
$\|\cdot\|_c$ in Eq.~\eqref{eq:normc} is the
largest expectation, in a qubit state, of the
corresponding linear combination of Bob's
measurements, when those measurements have overlap
$c$.

\textit{Step 3.-}
This upper bound is valid for every pair
$\vec\delta$. Minimizing over $\vec\delta$
therefore still gives an upper bound, and this
minimum is attained. The resulting minimum equals
the exact quantum bound at the given overlap.
For $|c|<1$, strong conic duality applies because
the fixed-overlap optimization has a strictly
feasible point. At $c=\pm1$, the corresponding
statement follows from linear-programming
duality.

\textit{Step 4.-}
Maximizing that exact overlap bound over
$c\in[-1,1]$ gives the exact quantum bound of
$\vec g$. The endpoints $c=\pm1$, at which Bob's
measurements commute, are included in the same
formula.
\qed

The full derivation of \eqref{eq:hQ} is given in
Appendix~\ref{app:dual}.

\section{Illustrating Theorem 1: Normal cones for a family of nonlocal extreme behaviours in $\mathcal{Q}_{2222}$}
\label{sec:idelta}

In this section, we illustrate our main
Theorem~\ref{thm:dualcone} and the algorithm in
Box~\ref{box:recipe} by constructing the normal cones
at a family of nonlocal extreme behaviours of high interest
for device-independent randomness.
In Ref.~\cite{WBC2022}, Wooltorton, Brown and Colbeck
introduced a one-parameter family of
Bell expressions $\mathcal I_\delta$ that tightly
bound the trade-off between nonlocality and
global randomness. At their quantum maxima,
these expressions certify two bits of global
randomness for the generation input pair
$(x,y)=(0,0)$, the maximum possible for two
binary outputs. We determine the normal cone at each of
the behaviours that maximize these expressions.
Specifically, for each $0<\delta\leq\pi/6$, the Bell expression of~\cite{WBC2022} is given as
\begin{equation}
\label{eq:Idelta}
\mathcal I_\delta
=E_{00}
+\frac1{\sin\delta}\,(E_{01}+E_{10})
-\frac1{\cos2\delta}\,E_{11},
\end{equation}
with a local maximum of $2 \operatorname{cosec}{\delta} + \operatorname{sec}{2\delta} - 1$. 
The unique maximizing behaviour within the quantum set $\mathcal{Q}_{2222}$ (achieving the quantum maximum $2\cot{\delta}+\tan{2\delta}$) is realized by the
maximally entangled state
$\bigl(|00\rangle+|11\rangle\bigr)/\sqrt2$, together
with the observables
\begin{equation}
\label{eq:Ideltaobs}
\begin{aligned}
A_0&=\sigma_z,&
A_1&=-\sin\delta\,\sigma_z+\cos\delta\,\sigma_x,\\
B_0&=\sigma_x,&
B_1&=\cos\delta\,\sigma_z-\sin\delta\,\sigma_x,
\end{aligned}
\end{equation}
where $\sigma_x$ and $\sigma_z$ are Pauli matrices.
The resulting behaviour is
\begin{equation}
\label{eq:pdelta}
\vec p_\delta
=\bigl(0,0,0,0,\,
0,\cos\delta,\cos\delta,-\sin2\delta\bigr),
\end{equation}
in the coefficient order
$(\alpha_0,\alpha_1,\beta_0,\beta_1,
E_{00},E_{01},E_{10},E_{11})$.
Bob's overlap is $c_0=-\sin\delta$.

The expression $\mathcal I_\delta$ is only one of
the Bell expressions maximized by $\vec p_\delta$.
The remaining freedom is the normal cone at this
behaviour. To explicitly find this normal cone, we normalize the quantum
value to one and write $\mathcal P_\delta$ for the
corresponding section, that is, the set of supporting
expressions with $\vec g\cdot\vec p_\delta=1$ given as
\begin{equation}
\label{eq:Pdelta}
\mathcal P_\delta
=\{\vec g\in N_{\mathcal Q}(\vec p_\delta):
\vec g\cdot\vec p_\delta=1\}.
\end{equation}
Let us denote $s=\sin\delta$, $c=\cos\delta$,
$K=\cos2\delta$, and $D=2c^3$. We then obtain the following theorem characterising the normal cones for this family of quantum extremal behaviors. 

\begin{widetext}
\begin{theorem}[Normal cones for the $I_\delta$ family]
\label{thm:idelta}
For every $0<\delta\leq\pi/6$, the set
$\mathcal P_\delta$ is the quadrilateral with vertices
\begin{equation}
\label{eq:Ideltaverts}
\begin{aligned}
\vec V_0
&=\frac1D\bigl(
(c-1)K,\;
s,\;
-sK,\;
(1-c)(2c+1),
sK,\;
K,\;
K,\;
-s
\bigr),\\
\vec V_1
&=\frac1D\bigl(
-sK,\;
(1-c)(2c+1),\;
(c-1)K,\;
s,
sK,\;
K,\;
K,\;
-s
\bigr),\\
\vec V_2
&=\frac1D\bigl(
sK,\;
(c-1)(2c+1),\;
(1-c)K,\;
-s,
sK,\;
K,\;
K,\;
-s
\bigr),\\
\vec V_3
&=\frac1D\bigl(
(1-c)K,\;
-s,\;
sK,\;
(c-1)(2c+1),
sK,\;
K,\;
K,\;
-s
\bigr),
\end{aligned}
\end{equation}
in the coefficient order
$(g_{\alpha_0},g_{\alpha_1},g_{\beta_0},g_{\beta_1},
g_{E_{00}},g_{E_{01}},g_{E_{10}},g_{E_{11}})$.
Every expression strictly inside the quadrilateral,
not on its boundary, has $\vec p_\delta$ as its unique
quantum maximizer.
\end{theorem}
\end{widetext}

That every expression in the interior of the
quadrilateral is uniquely maximized at
$\vec p_\delta$ follows from the known self-testing result for the $\mathcal I_\delta$ family from \cite{WBC2022}, i.e., any quantum
realization that attains the bound of
$\mathcal I_\delta$ is equivalent, up to local
isometries, to the state and measurements of
Eq.~\eqref{eq:Ideltaobs}. Note that the general construction
of Theorem~\ref{thm:dualcone} does not by itself
establish uniqueness at an arbitrary extremal
behaviour.

The vertex $\vec V_1$ is the exchange of the two
parties applied to $\vec V_0$. The vertices
$\vec V_2$ and $\vec V_3$ are obtained from
$\vec V_1$ and $\vec V_0$, respectively, by flipping
the signs of the four marginal coefficients.
The normalized expression $\mathcal I_\delta$ sits
at the centre of the quadrilateral. The quadrilateral
at $\delta=\pi/6$ is shown in Fig.~\ref{fig:idelta}.
The construction
and proof of Theorem~\ref{thm:idelta} are given in detail in Appendix~\ref{app:idelta}.

\begin{figure}[t]
\centering
\includegraphics[width=0.7\columnwidth]{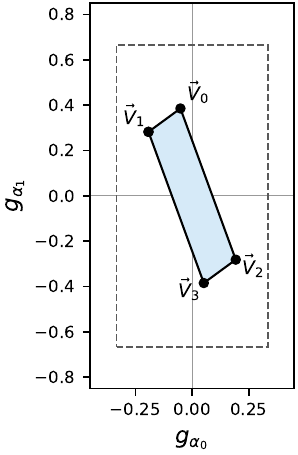}
\caption{Normalized supporting family at the two-bit
point $\delta=\pi/6$, in the plane of Alice's
marginal coefficients
$(g_{\alpha_0},g_{\alpha_1})$.
The four vertices are those of
Theorem~\ref{thm:idelta}, and the normalized
expression $\mathcal I_\delta$ sits at the origin
in this plane.
The dashed rectangle is the region of nonnegative
weights; the four local comparisons cut it to the
quadrilateral.}
\label{fig:idelta}
\end{figure}


While the above construction is for the family from
Ref.~\cite{WBC2022}, in Appendix~\ref{sec:example} we
illustrate the power and generality of
Theorem~\ref{thm:dualcone} by applying it to an
alternative extremal point.
Uniqueness of the interior supporting tests in that
example is then obtained from
Refs.~\cite{SunWangYu2026,LiuZingarofaloBB2026}.

\section{An application: Device-Independent Randomness with inefficient detectors}
\label{sec:lossy}

Thus far, we have seen the construction of the supporting family (the normal cone) consisting of all tests of one and
the same ideal extremal quantum behaviour. In a real experiment that
behaviour is not observed exactly, for instance because detectors
sometimes fail to click. Different supporting
expressions then give different scores, and can
certify different amounts of randomness. We now use
the complete family to choose an expression for an experiment incorporating detector loss.

\subsubsection{Detector loss}
Consider an honest experiment in which Alice and
Bob have independent detectors of equal efficiency
$\eta$. Whenever a detector fails to click, the
output $+1$ is recorded, and every trial is kept.
Randomness is generated from Alice's output for the
input $x=1$. This model describes only the honest
devices. The adversary is a general quantum adversary (holding general quantum side information) and is not assumed to suffer the
same loss, i.e., the randomness bound allows any quantum
realization compatible with the observed Bell score.

If the ideal behaviour is $\vec q$, the observed
behaviour $\vec p=L_\eta(\vec q)$ incorporating detector loss has outcome
biases
\begin{equation}
\label{eq:lossmarg}
\begin{aligned}
\alpha_x'&=\eta\alpha_x+1-\eta,\\
\beta_y'&=\eta\beta_y+1-\eta,
\end{aligned}
\end{equation}
and correlators
\begin{equation}
\label{eq:losscorr}
E_{xy}'=\eta^2E_{xy}
+\eta(1-\eta)(\alpha_x+\beta_y)
+(1-\eta)^2.
\end{equation}
These eight relations may be written in vector form as
\begin{equation}
\label{eq:lossmap}
L_\eta(\vec q)=\Lambda_\eta\vec q+\vec k_\eta,
\qquad
\mathcal Q_\eta=L_\eta(\mathcal Q).
\end{equation}
Here, the constant shift $\vec k_\eta$ is given as
\begin{equation}
\label{eq:lossshift}
\begin{split}
\vec k_\eta :=\bigl(&
1-\eta,\,1-\eta,\,1-\eta,\,1-\eta,\\
&(1-\eta)^2,\,(1-\eta)^2,\,(1-\eta)^2,\,(1-\eta)^2
\bigr).
\end{split}
\end{equation}
in the coefficient order
$(\alpha_0,\alpha_1,\beta_0,\beta_1,
E_{00},E_{01},E_{10},E_{11})$.
Analogously, the matrix $\Lambda_\eta$ produces the remaining,
$\eta$-dependent, part of the same relations, and is given as
\begin{equation}
\label{eq:lossLambda}
\begin{aligned}
(\Lambda_\eta\vec q)_{\alpha_x}&=\eta\alpha_x,\\
(\Lambda_\eta\vec q)_{\beta_y}&=\eta\beta_y,\\
(\Lambda_\eta\vec q)_{E_{xy}}
&=\eta^2E_{xy}+\eta(1-\eta)(\alpha_x+\beta_y).
\end{aligned}
\end{equation}
The set $\mathcal Q_\eta$ consists of all behaviours
that can be recorded from honest quantum devices of
efficiency $\eta$. A Bell expression $\vec g$,
evaluated on the recorded data, is equivalent to a
transformed expression on the ideal behaviour:
\begin{equation}
\label{eq:losspull}
\vec g\cdot L_\eta(\vec q)
=(\Lambda_\eta^{\mathsf T}\vec g)\cdot\vec q
+\vec g\cdot\vec k_\eta.
\end{equation}
For $\eta>0$ this relation can be inverted.
Consequently, if $\vec f$ supports the quantum set at
the ideal behaviour $\vec q$, the expression $\vec g$
defined by $\vec f=\Lambda_\eta^{\mathsf T}\vec g$ is
maximized at the recorded behaviour $L_\eta(\vec q)$
among all honest lossy behaviours in
$\mathcal Q_\eta$.
The available family of tests, adapted to detector loss, is
therefore the set of expressions $\vec g$ for which
\begin{equation}
\label{eq:losscone}
\Lambda_\eta^{\mathsf T}\vec g\in\mathcal C(\vec q).
\end{equation}
Note that this is only a change of the score used to test the
observed data. It does not undo the physical loss,
and it does not make the lossy behaviour a
maximizer over the unrestricted quantum set.

\subsubsection{Randomness certification}
We certify randomness from the observed Bell value
alone. Let $\Pguess(\vec g,\omega)$ denote the largest
probability with which an adversary holding
quantum side information can guess Alice's
output at input $x=1$, over all quantum
realizations whose average Bell value is
at least $\omega$. The corresponding
single-round min-entropy, which measures how
unpredictable that output remains~\cite{KonigRennerSchaffner2009}, is
\begin{equation}
\label{eq:hmin}
H_{\min}(A|X=1,E)
=-\log_2\Pguess(\vec g,\omega).
\end{equation}
For the honest experiment we take $\omega$ to be
the Bell value $\vec g\cdot\vec p$ of the observed
behaviour.

A single Bell value retains only one linear constraint
on the observed behaviour. If $\Pguess(\vec p)$
denotes the guessing probability when the entire
behaviour is known \cite{Bancal2014}, then
\begin{equation}
\label{eq:scorevsfull}
\Pguess(\vec g,\vec g\cdot\vec p)\geq\Pguess(\vec p).
\end{equation}
The randomness certified from the full statistics
is therefore always at least as large as that
certified from a single Bell
parameter~\cite{NietoSilleras2014,Bancal2014}.
However, estimation of the full behaviour, or of
several Bell expressions, leads to known difficulties in
parameter estimation~\cite{SeveralBell,RBHHW2016}.
The choice of $\vec g$ is therefore an attempt to retain, in
that one Bell value, the information most relevant to
randomness.

Our aim is, for any given practical detection efficiency $\eta \in [0.75,0.98]$, to construct Bell tests and two-qubit experiments that certify more single-round min-entropy from one Bell value
than any CHSH score can certify with the same
detector efficiency, even when the quantum
devices used for CHSH are optimized. To this end, we first search numerically over two-qubit honest
devices and over the loss-adapted family of
Eq.~\eqref{eq:losscone}. The devices and
expressions found in this way are then certified
independently of the search. Upper bounds on Eve's
guessing probability, together with nearly matching
explicit attacks, are given in
Appendix~\ref{app:lossy}
\cite{NPA2007,NPA2008}.
Appendix~\ref{app:own} explains the relation
between supporting expressions and the joint
choice of a device and a Bell score.

\subsubsection{Comparison with CHSH}
We compare the randomness certified from the
selected Bell score with that certified from a
CHSH score, at the same detector efficiency.
The CHSH expression is
$E_{00}+E_{01}+E_{10}-E_{11}$,
or any expression obtained from it by relabelling
inputs or outcomes. For a fair comparison we allow
every honest quantum device of efficiency $\eta$,
and every fixed rule that assigns a missing click
to an outcome $\pm1$. In both cases the adversary
is constrained only by the observed Bell value.

For a CHSH value $2\leq S\leq2\sqrt2$, the
min-entropy certified from that score is known to be
given exactly by the expression \cite{Pironio2010}
\begin{equation}
\label{eq:hminCHSH}
H_{\min}^{\mathrm{CHSH}}(S)
=-\log_2\!\left[
\frac{1+\sqrt{2-S^2/4}}{2}\right].
\end{equation}
Under the loss model above, the largest CHSH value
that can be recorded, and therefore the largest
min-entropy that a CHSH score can certify, is
determined in Ref.~\cite{Gigena2025}. Other sign
patterns and other assignments of the missing click
are bounded at the certified grid efficiencies; the
details are given in Appendix~\ref{app:lossy}.

Our benchmark optimizes the recorded CHSH score
over all honest quantum devices under the stated
loss model, including the partially entangled
strategies characterized in Ref.~\cite{Gigena2025}.
The comparison concerns single-round min-entropy
certified from one Bell score, not the critical
efficiency for detecting nonlocality.
We do not claim that the selected scores are
optimal among all Bell expressions or that they
outperform every other randomness protocol.

\begin{theorem}[Randomness beyond a CHSH score]
\label{thm:diqrng}
For every detector efficiency
$\eta\in[0.75,0.98]$, there is a two-qubit
experiment, allowing independent local click
discarding, and a Bell expression obtained from
the loss-adapted supporting construction at a
certified reference efficiency, such that the
single-round min-entropy certified from the
resulting Bell score is strictly larger than the
min-entropy certified from any CHSH score at
the same efficiency.
\end{theorem}

Table~\ref{tab:lossybrief} gives six reference
points. The selected scores improve on the
corresponding CHSH upper bounds by approximately
$9.2\%$--$18.0\%$.
The coefficients of these six selected Bell
expressions, together with the observed Bell
values that they attain, are listed in
Table~\ref{tab:lossyexprs} of Appendix~\ref{app:lossy}.

\begin{table}[t]
\centering
\caption{Single-round min-entropy in bits per
generation round, at the same detector efficiency.
The middle column is a certified lower bound on the
min-entropy obtained from the selected Bell score.
The last column is a certified upper bound on the
min-entropy obtained from any CHSH score.
All entries are rounded outwards.
Two-sided intervals for the selected scores
appear in Appendix~\ref{app:lossy}.}
\label{tab:lossybrief}
\small
\begin{tabular}{ccc}
\toprule
$\eta$ & Selected Bell score & Any CHSH score\\
\midrule
$0.75$ & $0.009845$ & $0.009013$\\
$0.80$ & $0.037859$ & $0.033445$\\
$0.85$ & $0.095104$ & $0.081930$\\
$0.90$ & $0.198626$ & $0.168313$\\
$0.95$ & $0.385565$ & $0.330204$\\
$0.98$ & $0.597367$ & $0.525646$\\
\bottomrule
\end{tabular}
\end{table}

The statement of Theorem~\ref{thm:diqrng} is
verified at $112$ certified efficiencies covering
the interval $[0.75,0.98]$. To pass from these
discrete values to every intermediate efficiency,
suppose the actual efficiency $\eta$ lies between
two neighbouring certified values
$\eta_i<\eta_{i+1}$. The detectors are then more
efficient than $\eta_i$, so Alice and Bob may,
independently on each side, treat a suitable
fraction of their clicks as missed detections
(still recording $+1$ and keeping every trial),
until the experiment reproduces the recorded
behaviour already certified at $\eta_i$.
The min-entropy certified at $\eta_i$ is therefore
a valid lower bound at the actual efficiency.
At the same time, a CHSH test at efficiency $\eta$
cannot certify more than a CHSH test at the higher
efficiency $\eta_{i+1}$, and we use a certified
upper bound on the latter. These two bounds remain
strictly separated throughout the grid, so the
advantage holds for every efficiency in the stated
interval. The six reference points of
Table~\ref{tab:lossybrief} and this continuum
coverage are reproduced from the public
archive~\cite{anc}.

The advantage concerns single-round certification
from one Bell score. It does not establish
optimality among all Bell expressions or an
advantage over certification using the full
observed behaviour. Finite-size performance
also depends on the input distribution and
statistical fluctuations, which are not included
in this comparison.

\section{Discussion}
\label{sec:discussion}

We have given an explicit construction of the
normal cone at every nonlocal extreme behaviour of the set of quantum correlations in the minimal Bell scenario
$\mathcal Q_{2222}$. The cone is obtained from four
tangent Bell expressions, a stationarity condition,
and sixteen local comparisons, and we proved that these conditions are sufficient.

We illustrated the construction on a family of nonlocal extreme behaviours from \cite{WBC2022} that has found application in device-independent randomness generation. Specifically, the normalized section of the normal cone is a quadrilateral at each of those behaviours; the full cone is the cone generated by that quadrilateral.

We also showed how the normal cone construction is useful even when an experiment is imperfect (the observed behaviour is not an extreme one). Different Bell expressions that all support the
same ideal extreme behaviour (belong to the normal cone of that behaviour) can certify different amounts
of randomness after detector loss. By choosing both
the device and the Bell expression, we obtain
throughout the detector efficiency interval
$\eta \in [0.75,0.98]$, a certified improvement over the
randomness that can be obtained from any CHSH score
at the same efficiency.
Related work has shown that incorporating zero-probability constraints can improve CHSH-based device-independent randomness expansion~\cite{ChenLiang2024}. Here we instead vary the supporting Bell expression at a given ideal behaviour under detector loss.

The normal cone construction is expected to find utility in a number of device-independent applications where $\mathcal{Q}_{2222}$ is an object of central interest. For an experimental implementation of a protocol such as for device-independent randomness generation, the natural next step is to include
finite statistics in that selection. A score
with a favorable single-round entropy bound may
also have large fluctuations, and the best
choice in that case can depend on the input probabilities and security parameters. The normal cone provides a structured set of Bell expressions
within which to address these questions.

Beyond the minimal Bell scenario, the same physical question remains: for which Bell tests does a given extreme quantum behaviour define the optimal quantum strategy, and which of these tests are most useful
under imperfections?
The exact construction given here relies on
the two-qubit structure of $\mathcal Q_{2222}$ and the explicit characterisation of this set discovered recently. Finding comparable constructions in larger
Bell scenarios is an interesting open direction.

\paragraph{Data availability.}
Ancillary code and certificate data are available at \url{https://github.com/subhramanyach-star/q2222-ancillary}, commit \texttt{29971d694306381f2510ffe614e8807e4a633082}. The accompanying \texttt{MANIFEST.md} identifies the inputs, verifiers, and expected outputs for the $I_\delta$ identities, the randomness bounds, the CHSH benchmark, and the efficiency-interval coverage~\cite{anc}.

\paragraph{Use of AI tools.}
Large language models (Grok~4 by xAI and GPT-6-Astra) assisted in developing the proof of Theorem~\ref{thm:dualcone} in Appendix~B through iterative discussion and feedback from the authors. The authors verified all steps and take full responsibility for the correctness of the results.

\paragraph{Note added.}
While preparing this manuscript following our initial work in \cite{RamanathanNDWU},
related work was posted by Sun, Wang, and Yu~\cite{SunWangYu2026}
and by Liu, Zingarofalo, Barizien, and Bancal~\cite{LiuZingarofaloBB2026}.
Sun, Wang, and Yu proved that every Bell functional with a strict quantum advantage has a unique maximizing quantum behaviour, established the biquadratic measurement-overlap determinant and its leading mixed coefficient (their Proposition 4), and analysed normal cones at nonlocal exposed points, including a complete cone certified in an explicit example by determinant-based positive-semidefiniteness arguments.
Liu et al.\ established uniqueness and self-testing for Bell inequalities with a quantum advantage, proved that normalized
supporting families at nonlocal extreme behaviours
have dimension at most two, and determined these
families along a partially entangled curve.
The same determinant identity also appears in their uniqueness argument.
In contrast, our main result is an explicit
necessary-and-sufficient construction of the
\emph{full normal cone} at every nonlocal extreme
behaviour. We also apply this construction
to device-independent randomness certification. 

\begin{acknowledgments}
We acknowledge support from the General Research Fund (GRF) Grant No.\ 17307925, and the Research Impact Fund (RIF) Grant No.\ R7035-21.
\end{acknowledgments}

\appendix
\newpage
\renewcommand{\thesubsubsection}{\thesection.\arabic{subsubsection}}

\numberwithin{equation}{section}

\section{An exact expression for the Quantum value of Bell expressions in the minimal scenario}
\label{app:dual}

This appendix proves Theorem~\ref{thm:dualmain} from the main text.
The derivation uses the NDWU characterization of
behaviours at a fixed measurement overlap \cite{RamanathanNDWU, SunZhouYu}. 
We begin by recalling the notation of Section~\ref{sec:notation}:
a behaviour is the vector $\vec p =(\alpha_0,\alpha_1,\beta_0,\beta_1,
E_{00},E_{01},E_{10},E_{11})$, and a Bell expression is 
$\vec g =(g_{\alpha_0}, g_{\alpha_1}, g_{\beta_0}, g_{\beta_1},
g_{E_{00}}, g_{E_{01}}, g_{E_{10}}, g_{E_{11}})$.

\subsubsection{Fixed-overlap behaviours}
For each Alice input $x$ and outcome $a$, define
\begin{equation}
\label{eq:tu}
t_{a|x}=1+a\alpha_x,
\qquad
\vec u_{a|x}
=
\begin{pmatrix}
\beta_0+aE_{x0}\\
\beta_1+aE_{x1}
\end{pmatrix}.
\end{equation}
Here $t_{a|x}/2=P(a|x)$ is the probability of the outcome $a$ given input $x$.
When $t_{a|x}>0$, the ratio $\vec u_{a|x}/t_{a|x}$ is the pair of
Bob's expectation values $w_{a|x,0}, w_{a|x,1}$ conditioned on Alice obtaining outcome
$a$ for input $x$, as in Eq.~\eqref{eq:w}.

For $c\in(-1,1)$, let
\[
G_c=
\begin{pmatrix}
1&c\\
c&1
\end{pmatrix}.
\]
The fixed-overlap set $K_c$ consists of those behaviours that
can be realized with Bob's measurement overlap held at the
value $c$. Explicitly,
\begin{equation}
\label{eq:Kc}
\begin{aligned}
K_c
=\bigl\{
\vec p:
{}&t_{a|x}\geq0,\\
&\vec u_{a|x}^{\mathsf T}G_c^{-1}\vec u_{a|x}
\leq t_{a|x}^{\,2}
\quad\text{for every }a,x
\bigr\}.
\end{aligned}
\end{equation}
Each inequality is the ellipse constraint of
Eq.~\eqref{eq:ndwu}, written for the unnormalized pair
$\vec u_{a|x}=t_{a|x}\vec w_{a|x}$. Equivalently, it may be
written as a bound on a vector length,
\begin{equation}
\label{eq:soc}
\left\|G_c^{-1/2}\vec u_{a|x}\right\|_2
\leq t_{a|x}.
\end{equation}
In particular, a zero-probability outcome, $t_{a|x}=0$, requires $\vec u_{a|x}=0$, and no conditional expectation needs to be assigned to that outcome.

The definitions in Eq.~\eqref{eq:tu} enforce
\begin{equation}
\label{eq:normids}
\sum_a t_{a|x}=2,
\qquad
\sum_a \vec u_{a|x}=2\vec\beta,
\qquad
\vec\beta=(\beta_0,\beta_1)^{\mathsf T}.
\end{equation}
The second identity holds for either input $x$. It expresses the requirement that Alice's two measurements yield the same unconditional averages on Bob's side.

At the degenerate overlaps $c=\pm1$, Eq.~\eqref{eq:Kc} is replaced by
\begin{equation}
\label{eq:cpm1}
u_{a|x,1}=c\,u_{a|x,0},
\qquad
|u_{a|x,0}|\leq t_{a|x},
\qquad
t_{a|x}\geq0.
\end{equation}
These are the parallel and antiparallel measurement limits, in
which Bob's two observables commute.

The NDWU characterization established by one of us in Ref.~\cite{RamanathanNDWU} gives
\begin{equation}
\label{eq:conv}
\mathcal Q
=
\operatorname{conv}
\left(
\bigcup_{c\in[-1,1]}K_c
\right).
\end{equation}
That is, every quantum behaviour is a mixture of behaviours
that each have a single fixed overlap. Consequently, for every
Bell coefficient vector $\vec g$, the optimal quantum value is given by
\begin{equation}
\label{eq:hKcmax}
h_{\mathcal Q}(\vec g)
=
\max_{c\in[-1,1]}h_{K_c}(\vec g),
\qquad
h_{K_c}(\vec g)=\max_{\vec p\in K_c}\vec g\cdot\vec p.
\end{equation}
Thus the remaining task is to calculate the Bell bound on each fixed-overlap set.

\subsubsection{An upper bound}
For any real pair $\vec v=(v_0,v_1)$, write
\begin{equation}
\label{eq:normcA}
\|\vec v\|_c=\sqrt{\vec v^{\mathsf T}G_c\vec v}.
\end{equation}
This is the same quantity as in Eq.~\eqref{eq:normc}.
For $|c|<1$, the Cauchy--Schwarz inequality and Eq.~\eqref{eq:soc} imply
\begin{equation}
\label{eq:CS}
\vec v\cdot\vec u_{a|x}
\leq
\|\vec v\|_c
\left\|G_c^{-1/2}\vec u_{a|x}\right\|_2
\leq
t_{a|x}\|\vec v\|_c.
\end{equation}
The same bound holds at $c=\pm1$: Eq.~\eqref{eq:cpm1} gives
\begin{equation}
\label{eq:CSpm1}
\vec v\cdot\vec u_{a|x}
=
(v_0+cv_1)u_{a|x,0}
\leq t_{a|x}|v_0+cv_1|
=
t_{a|x}\|\vec v\|_c.
\end{equation}

Write $\vec g_\beta=(g_{\beta_0},g_{\beta_1})$ and
$\vec g_{E_x}=(g_{E_{x0}},g_{E_{x1}})$ for the Bob and
correlator coefficients of $\vec g$. Let
$\vec\delta\in\mathbb R^2$ be an auxiliary vector, and define
\begin{equation}
\label{eq:Bx}
\vec\kappa_x=\frac{\vec g_\beta}2+(-1)^x\vec\delta.
\end{equation}
Since $\vec\kappa_0+\vec\kappa_1=\vec g_\beta$,
Eq.~\eqref{eq:normids} allows the Bell expression to be written as
\begin{equation}
\label{eq:gprewrite}
\vec g\cdot\vec p
=
\frac12\sum_{x,a}
\left[
a g_{\alpha_x}\,t_{a|x}
+
(\vec\kappa_x+a\vec g_{E_x})\cdot\vec u_{a|x}
\right].
\end{equation}
Applying Eq.~\eqref{eq:CS} or Eq.~\eqref{eq:CSpm1}, we obtain
\begin{equation}
\label{eq:gpbound}
\vec g\cdot\vec p
\leq
\frac12\sum_{x,a}
t_{a|x}
\left[
a g_{\alpha_x}+\|\vec\kappa_x+a\vec g_{E_x}\|_c
\right].
\end{equation}
For each $x$, the numbers $t_{a|x}/2$ form a probability distribution. Therefore
\begin{equation}
\label{eq:gpPhi}
\vec g\cdot\vec p\leq\Phi_c(\vec g,\vec\delta),
\end{equation}
with
\begin{equation}
\label{eq:Phi}
\begin{split}
\Phi_c(\vec g,\vec\delta)
&=
\sum_{x=0}^{1}
\max_{a=\pm1}
\\
&\qquad\left[
a g_{\alpha_x}+
\left\|
\frac{\vec g_\beta}2+(-1)^x\vec\delta+a\vec g_{E_x}
\right\|_c
\right].
\end{split}
\end{equation}
This bound holds for every $\vec p\in K_c$ and every $\vec\delta$. In particular,
\begin{equation}
\label{eq:hinf}
h_{K_c}(\vec g)\leq
\inf_{\vec\delta\in\mathbb R^2}\Phi_c(\vec g,\vec\delta).
\end{equation}
The freedom in $\vec\delta$ redistributes Bob's marginal coefficients between Alice's inputs without changing the Bell expression. Optimizing that redistribution tightens the bound.

\subsubsection{Equality}
For $|c|<1$, maximizing $\vec g\cdot\vec p$ over $K_c$ is a convex optimization problem whose constraints are the four length bounds in Eq.~\eqref{eq:soc}, together with the normalizations in Eq.~\eqref{eq:normids} and the equality of Bob's unconditional averages for the two inputs. The point
\[
\alpha_x=\beta_y=E_{xy}=0
\]
is strictly feasible: each $t_{a|x}=1$ and each $\vec u_{a|x}=0$. Because this interior point satisfies the constraints strictly, standard convex duality applies~\cite{BoydVandenberghe}: the primal optimum is finite, the dual optimum is attained, and the two values agree.

To identify the dual, introduce a two-component multiplier for the equality of Bob's unconditional averages. With the sign convention of Eq.~\eqref{eq:Bx}, this multiplier is $\vec\delta$. The remaining dual variables, denoted $\lambda_x$, enforce the normalization of Alice's outcome probabilities. The dual constraints are
\begin{equation}
\label{eq:dualcons}
\lambda_x
\geq
a g_{\alpha_x}+
\left\|
\frac{\vec g_\beta}2+(-1)^x\vec\delta+a\vec g_{E_x}
\right\|_c
\qquad
\text{for }a=\pm1.
\end{equation}
The dual objective is $\lambda_0+\lambda_1$. Eliminating $\lambda_x$ gives exactly Eq.~\eqref{eq:Phi}. Hence
\begin{equation}
\label{eq:hminPhi}
h_{K_c}(\vec g)
=
\min_{\vec\delta\in\mathbb R^2}\Phi_c(\vec g,\vec\delta).
\end{equation}

At $c=\pm1$, the constraints in Eq.~\eqref{eq:cpm1} define a linear program. Its feasible set is nonempty and compact, so linear-programming duality gives the same equality. Here the seminorm in Eq.~\eqref{eq:Phi} reduces to
\begin{equation}
\label{eq:seminorm}
\|\vec v\|_{\pm1}=|v_0\pm v_1|.
\end{equation}

Combining Eqs.~\eqref{eq:hKcmax} and~\eqref{eq:hminPhi} proves Theorem~\ref{thm:dualmain}:
\begin{equation}
\label{eq:A20}
h_{\mathcal Q}(\vec g)
=
\max_{c\in[-1,1]}
\min_{\vec\delta\in\mathbb R^2}
\Phi_c(\vec g,\vec\delta).
\end{equation}





\section{Proof of Theorem~\ref{thm:dualcone} - normal cones at nonlocal extremal behaviors}
\label{app:dualcone}

This appendix proves Theorem~\ref{thm:dualcone}.
We first characterize the supporting expressions
at fixed measurement overlap and derive the
necessary conditions.
We then prove sufficiency using the determinant
of the Bell operator.

Throughout, $\vec p$ is a fixed nonlocal extreme
behaviour. A variable behaviour is written
$\vec z$. The notation of Appendix~\ref{app:dual} is
used: for each input $x$ and outcome $a$,
\[
t_{a|x}=1+a\alpha_x,
\qquad
\vec u_{a|x}
=
\bigl(\beta_0+a E_{x0},\,\beta_1+a E_{x1}\bigr),
\]
so that $t_{a|x}/2=P(a|x)$ and, when
$t_{a|x}>0$, the ratio
$\vec w_{a|x}=\vec u_{a|x}/t_{a|x}$ is the pair of
Bob's conditional expectations, as in
Eq.~\eqref{eq:w} of the main text.

\subsubsection{Fixed-overlap normals}
\label{app:overlap-unique}

At a nonlocal behaviour, neither of Alice's two
measurements can have a deterministic outcome. If,
say, the outcome of $A_0$ were certain, one could
reproduce the behaviour by a local model: sample the
outcome of $A_1$ from its marginal, and respond on
Bob's side with the corresponding conditional
probabilities. Therefore
\begin{equation}
\label{eq:B2}
t_{a|x}>0
\qquad\text{for all }a,x.
\end{equation}
The four pairs $\vec w_{a|x}$ are therefore well
defined. They are computed from the eight
coordinates of $\vec p$ and do not depend on a
choice of realization.

By the NDWU characterization of
Ref.~\cite{RamanathanNDWU}, the quantum set is the
convex hull of the fixed-overlap sets $K_c$ of
Eq.~\eqref{eq:Kc}. In particular,
$K_c\subseteq\mathcal Q$ for every $c\in[-1,1]$.
For $|c|<1$, the inverse
\[
G_c^{-1}
=(1-c^2)^{-1}
\begin{pmatrix}
1&-c\\
-c&1
\end{pmatrix}
\]
converts each ellipse constraint
$\vec w_{a|x}^{\mathsf T}G_c^{-1}\vec w_{a|x}\le 1$
into a quadratic inequality in the overlap.
Precisely, writing
$\vec w_{a|x}=(w_{a|x,0},w_{a|x,1})$,
\[
\vec w_{a|x}^{\mathsf T}G_c^{-1}\vec w_{a|x}\le 1
\qquad\Longleftrightarrow\qquad
F_{a|x}(c)\le 0,
\]
where
\[
F_{a|x}(c)
=w_{a|x,0}^2+w_{a|x,1}^2+c^2-2c\,w_{a|x,0}w_{a|x,1}-1
\]
is the quadratic of Eq.~\eqref{eq:F} in the main
text. The equivalence uses $1-c^2>0$.

The target has a planar two-qubit
realization~\cite{Jordan,Masanes}, and therefore
belongs to some set $K_c$ with $c\in[-1,1]$. At the
endpoints $c=\pm1$, Bob's two measurements are
parallel or antiparallel and commute, and every
behaviour in $K_{\pm1}$ is local. Since $\vec p$ is
nonlocal, it belongs to some $K_c$ with
$c\in(-1,1)$.

The overlap is unique. This is
Lemma~\ref{lem:unique-overlap} of the main text; we
recall the argument for later use. Suppose
$\vec p\in K_c\cap K_d$ with $c\neq d$ and
$c,d\in(-1,1)$. Then $F_{a|x}(c)\le 0$ and
$F_{a|x}(d)\le 0$ for every $a,x$. The identity
\[
F_{a|x}\Bigl(\frac{c+d}{2}\Bigr)
=
\frac{F_{a|x}(c)+F_{a|x}(d)}{2}
-
\frac{(c-d)^2}{4}
\]
implies that $F_{a|x}(r)<0$ at the midpoint
$r=(c+d)/2$, for every pair. Thus all four pairs lie
strictly inside the ellipse at overlap $r$.

No nonlocal quantum behaviour is a vertex of the
no-signalling set~\cite{RTHH16}. One may therefore
vary $\vec p$ among nearby no-signalling behaviours
that keep every vanishing probability of $\vec p$
equal to zero. The conditions $t_{a|x}>0$ and
$F_{a|x}(r)<0$ are open, so a small enough
neighbourhood of this kind still satisfies the four
ellipse constraints at the overlap $r$, and is
therefore quantum. A short line segment through
$\vec p$ in that neighbourhood would express
$\vec p$ as a mixture of two different quantum
behaviours, contradicting extremality. Hence
$\vec p$ belongs to $K_{c_0}$ for a unique overlap
$c_0\in(-1,1)$.

At this unique overlap the four pairs lie on the
ellipse, not in its interior. This is
Lemma~\ref{lem:saturation} of the main text: a
nonlocal extreme behaviour admits a realization by a
pure two-qubit state and projective measurements,
with Bob's directions and the four conditional Bloch
vectors all in one plane; purity of the global state
together with Alice's projective measurements
implies that each conditional state is pure, so
$|\vec m_{a|x}|=1$; planarity then gives
\[
\vec w_{a|x}^{\mathsf T}G_{c_0}^{-1}\vec w_{a|x}
=|\vec m_{a|x}|^2
=1
\]
for every $a,x$. In the unnormalized variables of
Appendix~\ref{app:dual}, this is
\begin{equation}
\label{eq:B1}
\vec u_{a|x}^{\mathsf T}G_{c_0}^{-1}\vec u_{a|x}
=
t_{a|x}^{\,2}.
\end{equation}

As in the main text, set
\begin{equation}
\label{eq:B3}
\vec w_{a|x}=\frac{\vec u_{a|x}}{t_{a|x}},
\qquad
\vec v_{a|x}=G_{c_0}^{-1}\vec w_{a|x}.
\end{equation}
Saturation then gives
\begin{equation}
\label{eq:B4}
\vec v_{a|x}\cdot\vec w_{a|x}=1,
\qquad
\|\vec v_{a|x}\|_{c_0}=1.
\end{equation}

The four tangent expressions $\vec\gamma_{a|x}$ are
those of Eq.~\eqref{eq:gamma} in the main text. In
components,
\begin{equation}
\label{eq:B5}
\begin{aligned}
(\vec\gamma_{a|x})_{\alpha_{x'}}&=-a\,\delta_{xx'},\\
(\vec\gamma_{a|x})_{\beta_y}&=v_{a|x,y},
\qquad
(\vec\gamma_{a|x})_{E_{x'y}}
=
a\,\delta_{xx'}v_{a|x,y}.
\end{aligned}
\end{equation}

Each of these four expressions supports the
fixed-overlap set $K_{c_0}$ at $\vec p$. Indeed,
take any $\vec z\in K_{c_0}$. The ellipse constraint
at overlap $c_0$, together with
Eq.~\eqref{eq:B4}, implies
\[
\vec v_{a|x}\cdot \vec u_{a|x}(\vec z)
\leq t_{a|x}(\vec z).
\]
Hence
\begin{equation}
\label{eq:B6}
\vec\gamma_{a|x}\cdot \vec z
=
-a\alpha_x(\vec z)+\vec v_{a|x}\cdot \vec u_{a|x}(\vec z)
\leq1.
\end{equation}
At the target, equality holds:
\begin{equation}
\label{eq:B7}
\vec\gamma_{a|x}\cdot \vec p
=
-a\alpha_x+t_{a|x}=1.
\end{equation}
Thus each of the four expressions has value one at
$\vec p$, and does not exceed one on the set of
behaviours with Bob's overlap fixed at $c_0$.

The four tangent expressions generate the entire
normal cone of $K_{c_0}$ at $\vec p$:
\begin{equation}
\label{eq:B8}
N_{K_{c_0}}(\vec p)
=
\left\{
\sum_{a,x}\mu_{a|x}\vec\gamma_{a|x}:
\mu_{a|x}\geq0
\right\}.
\end{equation}

To see this, write the four ellipse constraints as
\begin{equation}
\label{eq:B9}
\Psi_{a|x}(\vec z)
=
\left\|G_{c_0}^{-1/2}\vec u_{a|x}(\vec z)\right\|_2
-t_{a|x}(\vec z)
\leq0.
\end{equation}
These four functions of $\vec z$ are convex. At
$\vec p$, each is differentiable, because the vector
$G_{c_0}^{-1/2}\vec u_{a|x}$ is nonzero: the four
pairs lie on the ellipse, not at the origin.
Differentiating gives
\begin{equation}
\label{eq:B10}
\nabla\Psi_{a|x}(\vec p)=\vec\gamma_{a|x}.
\end{equation}
The zero behaviour
$\alpha_x=\beta_y=E_{xy}=0$ satisfies all four
constraints strictly: each $t_{a|x}=1$ and each
$\vec u_{a|x}=0$. For a convex set defined by convex
inequalities that are differentiable at the point of
interest, and that can be satisfied strictly, every
supporting expression is a nonnegative combination
of the gradients of the active
constraints~\cite{BoydVandenberghe,Rockafellar}.
This is Eq.~\eqref{eq:B8}.

In particular, if
\begin{equation}
\label{eq:B11}
\vec g=\sum_{a,x}\mu_{a|x}\vec\gamma_{a|x},
\qquad \mu_{a|x}\geq0,
\end{equation}
then
\begin{equation}
\label{eq:B12}
\vec g\cdot \vec p=\sum_{a,x}\mu_{a|x},
\qquad
\vec g\cdot \vec z\leq \vec g\cdot \vec p
\quad\text{for every }\vec z\in K_{c_0}.
\end{equation}
Thus $\vec p$ maximizes $\vec g$ among all
behaviours with Bob's overlap equal to $c_0$. This
does not yet rule out a larger value at a different
overlap.

\subsubsection{Necessary conditions}

Suppose now that $\vec g$ supports the whole quantum
set at $\vec p$, that is,
$\vec g\in N_{\mathcal Q}(\vec p)$. Since
$K_{c_0}\subseteq\mathcal Q$, the same expression is
maximized by $\vec p$ on $K_{c_0}$.
Equation~\eqref{eq:B8} therefore gives the
representation~\eqref{eq:B11}.

A second condition is obtained by varying Bob's
overlap, as in the main text. From the unique
overlap $c_0$ and the four normals $\vec v_{a|x}$,
write a two-qubit realization of $\vec p$: choose
two unit vectors $\vec n_0$ and $\vec n_1$ in a
plane with $\vec n_0\cdot\vec n_1=c_0$, set
$\vec m_{a|x}=v_{a|x,0}\vec n_0+v_{a|x,1}\vec n_1$,
and form Bob's reduced state
$\rho_B=\sum_{a=\pm1}((1+a\alpha_x)/2)\,(I+\vec m_{a|x}\cdot\boldsymbol{\sigma})/2$,
independent of $x$. Let $\vec m_B$ be its Bloch
vector and $r=|\vec m_B|$. If $\vec m_B\neq0$, let
$|e_+\rangle$ and $|e_-\rangle$ be the qubit states
with Bloch vectors $\vec m_B/r$ and $-\vec m_B/r$;
if $\vec m_B=0$, take any orthonormal pair in the
plane. A shared state is
\[
|\psi\rangle
=\sqrt{\frac{1+r}{2}}\,|e_+\rangle_A|e_+\rangle_B
+\sqrt{\frac{1-r}{2}}\,|e_-\rangle_A|e_-\rangle_B,
\]
and, in the Schmidt basis used above, Alice's
measurement effects are
\[
\Pi_{a|x}
=\frac{1+a\alpha_x}{2}
\left(
\rho_B^{-1/2}\rho_{a|x}\rho_B^{-1/2}
\right)^{\mathsf T},
\]
with $A_x=\Pi_{+|x}-\Pi_{-|x}$ and
$\rho_{a|x}=(I+\vec m_{a|x}\cdot\boldsymbol{\sigma})/2$.
Here the transpose is taken in that basis,
with the resulting matrix acting on Alice's
qubit. Nonlocality implies that $\rho_B$ has
full rank, so the inverse is well defined.
For each input, these are two rank-one positive
operators summing to the identity, and hence
are orthogonal projectors.
Keep that state and Alice's measurements fixed, and
rotate $\vec n_0$ and $\vec n_1$ in the plane they
span, so that the overlap becomes a variable $c$.
This produces a smooth curve $\vec p(c)$ of quantum
behaviours with $\vec p(c_0)=\vec p$. Write
\begin{equation}
\label{eq:B13}
\dot{\vec p}=\left.\frac{d}{dc}\vec p(c)\right|_{c=c_0}.
\end{equation}
The function $f(c)=\vec g\cdot\vec p(c)$ has a
maximum at $c=c_0$, so
\begin{equation}
\label{eq:B14}
\vec g\cdot\dot{\vec p}=0.
\end{equation}

Finally, every local deterministic behaviour is
quantum, so
\begin{equation}
\label{eq:B15}
\vec g\cdot\vec d\leq \vec g\cdot \vec p
\end{equation}
for each of the sixteen local deterministic behaviours
$\vec d$ of Eq.~\eqref{eq:detvec}, or equivalently
\begin{equation}
\label{eq:B16}
h_{\mathrm L}(\vec g)\leq \vec g\cdot \vec p.
\end{equation}

These three conditions are the definition of the
candidate cone. We have therefore shown
\begin{equation}
\label{eq:B17}
N_{\mathcal Q}(\vec p)\subseteq\mathcal C(\vec p),
\end{equation}
where
\begin{equation}
\label{eq:B18}
\begin{aligned}
\mathcal C(\vec p)
=\bigl\{
\vec g=\sum_{a,x}\mu_{a|x}\vec\gamma_{a|x}:{}
&\mu_{a|x}\geq0,\\
&\vec g\cdot\dot{\vec p}=0,\\
&h_{\mathrm L}(\vec g)\leq \vec g\cdot \vec p
\bigr\}.
\end{aligned}
\end{equation}

In terms of the four weights alone, the same
conditions read
\[
\mu_{a|x}\geq0,
\qquad
\sum_{a,x}\mu_{a|x}
\bigl(\vec\gamma_{a|x}\cdot\dot{\vec p}\bigr)=0,
\]
and
\begin{equation}
\label{eq:B19}
\sum_{a,x}\mu_{a|x}
\bigl(\vec\gamma_{a|x}\cdot\vec d-1\bigr)\leq0
\end{equation}
for each of the sixteen local deterministic behaviours
$\vec d$. The candidate cone is the set of all
expressions $\vec g=\sum_{a,x}\mu_{a|x}\vec\gamma_{a|x}$
whose weights satisfy these linear conditions.

\subsubsection{The Bell operator}
\label{app:suff-start}

It remains to prove the converse: every
$\vec g\in\mathcal C(\vec p)$ supports the whole
quantum set at $\vec p$.

The three conditions already ensure that $\vec p$
maximizes $\vec g$ among behaviours with overlap
$c_0$, that the value is stationary when Bob's
overlap is varied, and that no local deterministic
behaviour exceeds the target value. What must still
be shown is that no quantum behaviour at a different
overlap can exceed it either.

Write
\begin{equation}
\label{eq:B20}
f_{\vec g}(c)=h_{K_c}(\vec g)
=
\min_{\vec\delta}\Phi_c(\vec g,\vec\delta)
\end{equation}
for the exact Bell bound on the set of behaviours
with Bob's overlap fixed at $c$, as given by
Theorem~\ref{thm:dualmain} and
Appendix~\ref{app:dual}. For every
$\vec g\in\mathcal C(\vec p)$,
Eq.~\eqref{eq:B12} gives
\begin{equation}
\label{eq:B21}
f_{\vec g}(c_0)=\vec g\cdot \vec p.
\end{equation}
The claim is
\begin{equation}
\label{eq:B22}
f_{\vec g}(c)\leq f_{\vec g}(c_0)
\qquad\text{for every }c\in[-1,1],
\end{equation}
and therefore $h_{\mathcal Q}(\vec g)=\vec g\cdot\vec p$.

\label{app:suff-setup}
\allowdisplaybreaks

After local unitaries, every nonlocal extreme
behaviour of $\mathcal Q_{2222}$ admits a planar
two-qubit realization~\cite{Masanes}: the state may
be taken as
\begin{equation}
\label{eq:suff-psi}
\lvert\psi_\theta\rangle=\cos\theta\,\lvert 00\rangle+\sin\theta\,\lvert 11\rangle,
\qquad 0<\theta\le\pi/4,
\end{equation}
and the four observables may be taken in a single
plane,
\begin{align*}
A_x&=\cos\varphi_x\,\sigma_z+\sin\varphi_x\,\sigma_x,\\
B_y&=\cos\chi_y\,\sigma_z+\sin\chi_y\,\sigma_x.
\end{align*}
Bob's overlap at the target is
$c_0=\cos(\chi_1-\chi_0)$. We assume throughout that
Bob's two directions are not parallel or
antiparallel, so that $\sin(\chi_0-\chi_1)\neq 0$
and $G_{c_0}$ is invertible, and that the behaviour
is nonlocal, so that none of Alice's outcomes is
deterministic.

The conditional Bloch vector of Bob's state after
Alice obtains outcome $a$ on input $x$, written in
the $(\sigma_x,\sigma_z)$ coordinates, is
\[
\vec m_{a|x}
=
\frac{1}{1+a\cos 2\theta\cos\varphi_x}
\begin{pmatrix}
a\sin 2\theta\sin\varphi_x\\
\cos 2\theta+a\cos\varphi_x
\end{pmatrix}.
\]
It has unit length, as required by
Lemma~\ref{lem:saturation}. The corresponding
outcome weight is
$t_{a|x}=1+a\cos 2\theta\cos\varphi_x$.

The candidate cone consists of the nonnegative
combinations
\begin{equation}
\label{eq:suff-gmu}
\vec g=\sum_{a,x}\mu_{a|x}\vec\gamma_{a|x},
\qquad
\mu_{a|x}\ge 0,
\end{equation}
that satisfy $\vec g\cdot\dot{\vec p}=0$ and
$h_{\mathrm L}(\vec g)\le\vec g\cdot\vec p$.
Necessity has already been proved. The task is the
reverse inclusion.

When $\sum_{a,x}\mu_{a|x}=1$, we say that the
combination is normalized; then
$\vec g\cdot\vec p=1$ by Eq.~\eqref{eq:B12}.

For later calculations it is convenient to record
the derivative of the behaviour with respect to one
of Bob's measurement angles. Write
\begin{equation}
\label{eq:suff-m}
\begin{aligned}
m_{a|x}
&:=\vec\gamma_{a|x}\cdot
\frac{\partial\vec p}{\partial\chi_0}\\
&=\frac{t_{a|x}}{D}
\bigl(\vec m_{a|x}\cdot \vec n_0^\perp\bigr)
\bigl(\vec m_{a|x}\cdot \vec n_1^\perp\bigr),
\end{aligned}
\end{equation}
where $D=\sin(\chi_0-\chi_1)$ and
$\vec n_y^\perp=(\cos\chi_y,-\sin\chi_y)$ is the
direction in the plane orthogonal to Bob's
measurement for input $y$, written in the
$(\sigma_x,\sigma_z)$ coordinates. Since
$\partial c/\partial\chi_0=-D\neq 0$, one has
\[
m_{a|x}
=-D\,\vec\gamma_{a|x}\cdot\dot{\vec p}.
\]
The stationarity condition
$\vec g\cdot\dot{\vec p}=0$ is therefore equivalent
to $\sum_{a,x}\mu_{a|x}m_{a|x}=0$. The derivative
with respect to $\chi_1$ is $-m_{a|x}$.

\begin{lemma}[The target state is an eigenstate]
\label{lem:kernel}
Let $H(\vec g)$ be the Bell operator formed from the
expression $\vec g$ and the observables of the
target realization: that is, the operator on the
two-qubit space whose expectation in a state is the
Bell value of $\vec g$ in the corresponding
behaviour. Then, for each tangent expression,
\[
H(\vec\gamma_{a|x})\lvert\psi_\theta\rangle
=\lvert\psi_\theta\rangle.
\]
Consequently, for
$\vec g=\sum_{a,x}\mu_{a|x}\vec\gamma_{a|x}$,
\[
H(\vec g)\lvert\psi_\theta\rangle
=\left(\sum_{a,x}\mu_{a|x}\right)\lvert\psi_\theta\rangle,
\qquad
\vec g\cdot \vec p=\sum_{a,x}\mu_{a|x}.
\]
In particular, if the combination is normalized,
both the eigenvalue and the target value are equal
to one.
\end{lemma}

\begin{proof}
Let $\sigma^{a|x}=\vec m_{a|x}\cdot\boldsymbol{\sigma}$ be
the observable that measures Bob's qubit along the
conditional Bloch vector. By construction of
$\vec\gamma_{a|x}$,
\[
H(\vec\gamma_{a|x})
=-aA_x\otimes I+(I+aA_x)\otimes\sigma^{a|x}.
\]
Bob's conditional state for outcome $a$ is the
$+1$ eigenstate of $\sigma^{a|x}$. Therefore
\[
\bigl[(I+aA_x)\otimes\sigma^{a|x}\bigr]
\lvert\psi_\theta\rangle
=
\bigl[(I+aA_x)\otimes I\bigr]
\lvert\psi_\theta\rangle.
\]
The first identity follows, and the remaining
statements follow by linearity.
\end{proof}

Combinations of the four tangents are automatically
stationary under variations that leave Bob's overlap
unchanged. Indeed, each tangent expression is
already maximized at $\vec p$ among behaviours with
the same Bob overlap, so a variation of the state,
of Alice's measurements, or of a common rotation of
Bob's two directions (which does not change the
overlap) cannot increase the value at first order.
By linearity, the same holds for every nonnegative
combination. The only remaining stationarity
condition is the change of overlap,
$\sum_{a,x}\mu_{a|x}m_{a|x}=0$.

The candidate cone at a nonlocal extreme point is
not reduced to zero. The quantum set is compact and
convex, and has nonempty interior in the
no-signalling space. Every extreme point therefore
lies on the boundary and admits a nonzero supporting
expression. By the necessity already proved, that
expression lies in $\mathcal C(\vec p)$. Because
$\vec g\cdot\vec p=\sum_{a,x}\mu_{a|x}$ for every
combination of the four tangents, it follows that
$\mathcal C(\vec p)$ contains a normalized
expression.

To study the Bell value at other overlaps, it is
convenient to rotate the local axes so that
$A_0=B_0=\sigma_z$, while $A_1$ and $B_1$ remain in
the $\sigma_x$--$\sigma_z$ plane. This is always
possible by a local rotation in the plane of the
measurements. In these axes,
\[
A_1=t\sigma_z+s_A\sigma_x,
\qquad
B_1=c\sigma_z+s_B\sigma_x,
\]
where $t\in[-1,1]$ is the overlap of Alice's two
directions, $c\in[-1,1]$ is the overlap of Bob's two
directions, and $s_A^2=1-t^2$, $s_B^2=1-c^2$. The
eight coefficients of $\vec g$ are unchanged: only
the operator representatives are rotated. The
possible pairs of overlaps $(t,c)$ fill the square
$[-1,1]\times[-1,1]$. On the boundary of this
square, at least one party has commuting
measurements: $t=\pm1$ means $A_1=\pm A_0$, and
$c=\pm1$ means $B_1=\pm B_0$. The target
measurements correspond to an interior point
\[
t_*=\cos(\varphi_1-\varphi_0),
\qquad
c_*=\cos(\chi_1-\chi_0),
\]
which lies in the open square whenever both parties
have non-commuting planar settings. Here $c_*$ is
the same overlap as $c_0$.

Let $H(t,c)$ be the Bell operator of the same
expression $\vec g$ in these axes, and write
$G(t,c)=I_4-H(t,c)$ and
$\Delta(t,c)=\det G(t,c)$. The Bell value in a
two-qubit state $\lvert\psi\rangle$ is
$\langle\psi\rvert H\lvert\psi\rangle$. If $I_4-H$
is positive semidefinite, then
$\langle\psi\rvert H\lvert\psi\rangle\le 1$ for
every state, so no two-qubit realization at those
overlaps can exceed the value one. The determinant
$\Delta$ will be used as a test for this, once the
other eigenvalue conditions are under control, in
Lemma~\ref{lem:upgrade} below. After the sines have
been replaced by $s_A^2=1-t^2$ and $s_B^2=1-c^2$,
the determinant does not depend on the signs of
those sines.

To compute $\Delta$, work in orthonormal coordinates
on the two measurement planes and write
\[
H=\mathbf A\cdot\boldsymbol{\sigma}\otimes I
+I\otimes\mathbf B\cdot\boldsymbol{\sigma}
+\sum_{i,j}T_{ij}\sigma_i\otimes\sigma_j,
\]
where
\[
\mathbf A=N\vec g_\alpha,
\qquad
\mathbf B=M\vec g_\beta,
\qquad
T=N\mathsf E M^{\mathsf T}.
\]
Here the columns of $N$ and $M$ are the measurement
directions of Alice and of Bob, respectively, and
\[
\vec g_\alpha=\begin{pmatrix}g_{\alpha_0}\\g_{\alpha_1}\end{pmatrix},
\qquad
\vec g_\beta=\begin{pmatrix}g_{\beta_0}\\g_{\beta_1}\end{pmatrix},
\qquad
\mathsf E=
\begin{pmatrix}
g_{E_{00}}&g_{E_{01}}\\
g_{E_{10}}&g_{E_{11}}
\end{pmatrix}
\]
is the matrix of correlator coefficients. Thus
$N^{\mathsf T}N=J_t$ and $M^{\mathsf T}M=J_c$, with
\[
J_t=\begin{pmatrix}1&t\\t&1\end{pmatrix},
\qquad
J_c=\begin{pmatrix}1&c\\c&1\end{pmatrix}.
\]
In particular,
$(\det T)^2=(\det\mathsf E)^2(1-t^2)(1-c^2)$.
Define
\[
R=\vec g_\alpha^{\mathsf T}J_t\vec g_\alpha
+\vec g_\beta^{\mathsf T}J_c\vec g_\beta
+\operatorname{tr}(\mathsf E^{\mathsf T}J_t\mathsf E J_c),
\]
\[
Q=\vec g_\alpha^{\mathsf T}J_t\mathsf E J_c\vec g_\beta,
\]
and
\[
\begin{aligned}
W={}&
\vec g_\beta^{\mathsf T}J_c\mathsf E^{\mathsf T}J_t\mathsf E J_c\vec g_\beta\\
&+\vec g_\alpha^{\mathsf T}J_t\mathsf E J_c\mathsf E^{\mathsf T}J_t\vec g_\alpha\\
&+(\vec g_\alpha^{\mathsf T}J_t\vec g_\alpha)(\vec g_\beta^{\mathsf T}J_c\vec g_\beta)\\
&+(\det\mathsf E)^2(1-t^2)(1-c^2).
\end{aligned}
\]
Pauli multiplication gives the traces
\begin{align*}
\operatorname{tr}H^2&=4R,\\
\operatorname{tr}H^3&=24Q,\\
\operatorname{tr}H^4&=4R^2+16W.
\end{align*}
Since $\operatorname{tr}H=0$, the coefficients of
the characteristic polynomial are determined by
these traces, and
\begin{equation*}
\begin{aligned}
\det(I-H)
&=
1-\tfrac12\operatorname{tr}H^2
-\tfrac13\operatorname{tr}H^3\\
&\qquad
+\tfrac18(\operatorname{tr}H^2)^2
-\tfrac14\operatorname{tr}H^4.
\end{aligned}
\end{equation*}
Therefore
\[
\Delta=1-2R-8Q+R^2-4W.
\]

Both $R$ and $Q$ are bilinear in $(t,c)$. The first
term of $W$ has degree at most one in $t$ and two in
$c$; the second has degree at most two in $t$ and
one in $c$; the remaining terms have degree at most
two in each variable. Thus $\Delta$ is a polynomial
of degree at most two in $t$ and at most two in $c$.

The coefficient of $tc$ in $R$ is
$2(g_{E_{00}}g_{E_{11}}+g_{E_{01}}g_{E_{10}})$.
Only $R^2$ and the final term of $W$ contribute to
$t^2c^2$, so
\begin{equation}
\label{eq:suff-lead-tc}
\begin{aligned}
[t^2c^2]\,\Delta
&=4(g_{E_{00}}g_{E_{11}}+g_{E_{01}}g_{E_{10}})^2\\
&\qquad-4(g_{E_{00}}g_{E_{11}}-g_{E_{01}}g_{E_{10}})^2\\
&=16\prod_{x,y}g_{E_{xy}}.
\end{aligned}
\end{equation}
(This biquadratic determinant and its leading mixed coefficient were established in Proposition 4 of Ref.~\cite{SunWangYu2026}; see also the uniqueness argument in Lemma 13 of Ref.~\cite{LiuZingarofaloBB2026}. Here we use this structure to establish global quantum feasibility from our candidate conditions.)

For a normalized combination in the candidate cone,
Lemma~\ref{lem:kernel} gives
$G(t_*,c_*)\lvert\psi_*\rangle=0$, and therefore
$\Delta(t_*,c_*)=0$. Moreover, the Bell value is
stationary at the target under a change of either
overlap: the variations that preserve the overlap
cannot increase the value at first order, and the
imposed condition $\vec g\cdot\dot{\vec p}=0$ takes
care of a change of Bob's overlap. Changing to the
axes above does not alter this conclusion:
derivatives of the transformed state contribute
nothing, because that state is an eigenstate of
eigenvalue one. Write $G_*=G(t_*,c_*)$. For either
overlap coordinate $\xi\in\{t,c\}$,
\[
\partial_\xi\Delta
=\operatorname{tr}\!\bigl(\operatorname{adj}(G)\,\partial_\xi G\bigr).
\]
If the kernel of $G_*$ has dimension at least two,
then the adjugate $\operatorname{adj}(G_*)$ vanishes.
If the kernel is one-dimensional, then
$\operatorname{adj}(G_*)=\kappa\lvert\psi_*\rangle\langle\psi_*\rvert$
for some real $\kappa$, and consequently
\[
\partial_\xi\Delta\big|_*
=-\kappa\bigl\langle\psi_*\bigr\rvert\partial_\xi H\bigr\lvert\psi_*\bigr\rangle
=0.
\]
Thus $\nabla\Delta(t_*,c_*)=0$: the target is an
interior point at which $\Delta$ vanishes together
with its first derivatives.

\subsubsection{Sufficiency}
\label{app:suff-line}

First suppose that $\prod_{x,y}g_{E_{xy}}\leq0$.
In this case the
leading coefficient of $\Delta$ is not positive. We
show that $\Delta$ cannot become negative anywhere
in the square of overlaps, and therefore that no
quantum realization can exceed the target value.

Determinant-based positive-semidefiniteness arguments
were also used for global quantum-bound certification
in Appendix D of Ref.~\cite{SunWangYu2026}.

\begin{lemma}[Positivity criterion]
\label{lem:upgrade}
Let $\vec g$ be an arbitrary Bell expression with
$h_{\mathrm L}(\vec g)\le 1$. Then, throughout the
square of overlaps,
\[
I_4-H(t,c)\succeq 0
\quad\text{if and only if}\quad
\Delta(t,c)\ge 0.
\]
\end{lemma}

\begin{proof}
Write $G=I_4-H$. For any Bell expression one has
$\operatorname{tr} H=0$, hence $\operatorname{tr} G=4$.
The elementary symmetric sums of the eigenvalues of
$G$ are then
\begin{align*}
\sigma_2(G)&=6-\tfrac12\operatorname{tr}(H^2),\\
\sigma_3(G)&=4-\operatorname{tr}(H^2)-\tfrac13\operatorname{tr}(H^3).
\end{align*}
In the axes above, after replacing the sines by
$s_A^2=1-t^2$ and $s_B^2=1-c^2$, both traces are
bilinear in $(t,c)$:
\begin{align*}
\operatorname{tr}(H^2)
&=4\bigl(g_{\alpha_0}^2+g_{\alpha_1}^2+g_{\beta_0}^2+g_{\beta_1}^2\\
&\qquad+g_{E_{00}}^2+g_{E_{01}}^2+g_{E_{10}}^2+g_{E_{11}}^2\bigr)\\
&\quad+8t\bigl(g_{\alpha_0} g_{\alpha_1}+g_{E_{00}}g_{E_{10}}+g_{E_{01}}g_{E_{11}}\bigr)\\
&\quad+8c\bigl(g_{\beta_0} g_{\beta_1}+g_{E_{00}}g_{E_{01}}+g_{E_{10}}g_{E_{11}}\bigr)\\
&\quad+8tc\bigl(g_{E_{00}}g_{E_{11}}+g_{E_{01}}g_{E_{10}}\bigr),
\end{align*}
and
\begin{align*}
\operatorname{tr}(H^3)
&=24\bigl(g_{\alpha_0} g_{\beta_0} g_{E_{00}}+g_{\alpha_0} g_{\beta_1} g_{E_{01}}\\
&\qquad+g_{\alpha_1} g_{\beta_0} g_{E_{10}}+g_{\alpha_1} g_{\beta_1} g_{E_{11}}\bigr)\\
&\quad+24t\bigl(g_{\alpha_0} g_{\beta_0} g_{E_{10}}+g_{\alpha_0} g_{\beta_1} g_{E_{11}}\\
&\qquad+g_{\alpha_1} g_{\beta_0} g_{E_{00}}+g_{\alpha_1} g_{\beta_1} g_{E_{01}}\bigr)\\
&\quad+24c\bigl(g_{\alpha_0} g_{\beta_0} g_{E_{01}}+g_{\alpha_0} g_{\beta_1} g_{E_{00}}\\
&\qquad+g_{\alpha_1} g_{\beta_0} g_{E_{11}}+g_{\alpha_1} g_{\beta_1} g_{E_{10}}\bigr)\\
&\quad+24tc\bigl(g_{\alpha_0} g_{\beta_0} g_{E_{11}}+g_{\alpha_0} g_{\beta_1} g_{E_{10}}\\
&\qquad+g_{\alpha_1} g_{\beta_0} g_{E_{01}}+g_{\alpha_1} g_{\beta_1} g_{E_{00}}\bigr).
\end{align*}
(The apparent quadratic terms in $t$ or $c$ cancel
against $s_A^2=1-t^2$ and $s_B^2=1-c^2$.)
Consequently $\sigma_2$ and $\sigma_3$ are bilinear
on the square $t,c\in[-1,1]$. A bilinear function on
a square attains its extrema at the four corners.
Those corners are points at which at least one party
has commuting measurements, so the resulting
behaviour is local, and $h_{\mathrm L}(\vec g)\le 1$
forces $G\succeq 0$. Hence $\sigma_2\ge 0$ and
$\sigma_3\ge 0$ at the corners, and therefore
throughout the square.

Suppose also that $\Delta=\det G\ge0$.
The characteristic polynomial evaluated at
$-\lambda$ can be written as
\[
\det(\lambda I+G)
=
\lambda^4+4\lambda^3
+\sigma_2(G)\lambda^2
+\sigma_3(G)\lambda+\Delta.
\]
This expression is strictly positive for every
$\lambda>0$. If $G$ had a negative eigenvalue
$-\lambda$, with $\lambda>0$, the determinant
would vanish, a contradiction. Hence $G\succeq0$.
The converse follows immediately from
positive semidefiniteness.
\end{proof}

The same local comparison supplies nonnegativity of
the determinant on the boundary. If
$h_{\mathrm L}(\vec g)\le 1$, then on each edge of
the square at least one party has commuting
measurements, the behaviour is local, so $G\succeq 0$
and therefore $\Delta\ge 0$ on the boundary.

\begin{lemma}[Polynomial lemma]
\label{lem:line}
Let $\Delta(t,c)$ be a real polynomial of degree at
most two in each variable, on the closed square
$R=[-1,1]\times[-1,1]$, and let
$(t_*,c_*)$ be an interior point of $R$. Suppose
the leading coefficient satisfies
$A:=[t^2c^2]\,\Delta\le 0$, that $\Delta$ and
$\nabla\Delta$ vanish at $(t_*,c_*)$, and that
$\Delta\ge 0$ on the boundary of $R$. Then
$\Delta\ge 0$ on the whole square.
\end{lemma}

\begin{proof}
Fix a direction $(u,v)\in\mathbb R^2$, not both
zero, and restrict to the line
$t=t_*+\sigma u$, $c=c_*+\sigma v$. The restriction
$F(\sigma):=\Delta(t_*+\sigma u,\,c_*+\sigma v)$ is
a polynomial of degree at most four. The only
degree-four monomial of $\Delta$ is $A t^2 c^2$, and
the change of variables contributes
$[\sigma^4]F=A u^2 v^2$, independently of the base
point: every other monomial of $\Delta$ has total
degree at most three in $(t,c)$, hence at most
three in $\sigma$. The conditions
$F(0)=\Delta(t_*,c_*)=0$ and
$F'(0)=\nabla\Delta(t_*,c_*)\cdot(u,v)=0$ therefore
give $F(\sigma)=\sigma^2 q(\sigma)$, with $q$ a
polynomial of degree at most two and
\[
q''(\sigma)=2A u^2 v^2.
\]
If $A\le 0$, then $q''\le 0$, so $q$ is concave.
(If $u=0$ or $v=0$, the line is parallel to one
axis, $q''=0$, and $q$ is affine, hence still
concave.) A linear term in $q$, coming from
coefficients of $\Delta$ that are odd in $t$ or in
$c$, does not enter $q''$.

The square $R$ is convex. For any point $P\in R$,
the intersection of $R$ with the line through
$(t_*,c_*)$ and $P$ is a closed segment whose
endpoints lie on the boundary and which contains
both the target and $P$. Parametrizing that segment
as above, the parameter interval is
$[\sigma_-,\sigma_+]$ with $\sigma_-<0<\sigma_+$
whenever the target is interior and
$P\neq(t_*,c_*)$. Nonnegativity on the boundary
gives $F(\sigma_\pm)\ge 0$. Since $\sigma_\pm\neq 0$,
one has $q(\sigma_\pm)=F(\sigma_\pm)/\sigma_\pm^2\ge 0$.
A concave function on a compact interval attains
its minimum at an endpoint, so $q\ge 0$ on
$[\sigma_-,\sigma_+]$, hence $F\ge 0$ on the
segment, and in particular $\Delta(P)\ge 0$. At the
target itself $\Delta=0$.
\end{proof}

\begin{proposition}[Sufficiency for nonpositive correlator product]
\label{prop:first-order}
Let $\vec p$ be generated by the pure planar
two-qubit realization of Eq.~\eqref{eq:suff-psi},
with non-commuting measurements on each side, and
let $\vec\gamma_{a|x}$ be its four tangent
expressions. Let
$\vec g=\sum_{a,x}\mu_{a|x}\vec\gamma_{a|x}$,
where $\mu_{a|x}\geq0$ and
$\sum_{a,x}\mu_{a|x}=1$,
with $\vec g\cdot\dot{\vec p}=0$,
$h_{\mathrm L}(\vec g)\le 1$, and
$g_{E_{00}}g_{E_{01}}g_{E_{10}}g_{E_{11}}\le 0$.
Then $h_{\mathcal Q}(\vec g)=1$.
\end{proposition}

\begin{proof}
In the axes above, the target $(t_*,c_*)$ lies in
the open square; $\Delta$ is of degree at most two
in each overlap, with leading coefficient
$16\prod_{x,y}g_{E_{xy}}\le 0$; $\Delta$ and
$\nabla\Delta$ vanish at the target by
Lemma~\ref{lem:kernel} and by the stationarity
already noted; and $\Delta\ge 0$ on the boundary of
the square by $h_{\mathrm L}(\vec g)\le 1$.
Lemma~\ref{lem:line} gives $\Delta\ge 0$ on the
square, and Lemma~\ref{lem:upgrade} gives
$I_4-H\succeq 0$ throughout. Thus no planar
two-qubit realization can produce a Bell value
larger than $1$. The same operator inequality holds
for every state on each two-qubit block, not only
for planar real states. Jordan's lemma decomposes a
general projective realization into blocks of local
dimension at most two. The bound holds on each block
and hence on their mixture. Binary measurements that
are not projective are covered by dilation to a
projective measurement on a larger space, which
again decomposes into two-qubit blocks. Thus
$h_{\mathcal Q}(\vec g)=1$. Since
$\vec g\cdot \vec p=1$ by Lemma~\ref{lem:kernel},
the target attains the bound.
\end{proof}

\label{app:suff-cell}

It remains to exclude $\prod_{x,y}g_{E_{xy}}>0$.
This case can be excluded directly from
positivity of the probabilities. No stationarity or
tangent representation is needed for this step.

\begin{lemma}[Local domination]
\label{lem:local-domination}
Let $\vec p$ be a nonlocal binary-output
no-signalling behaviour in this scenario, and let
$\vec g$ be a Bell expression. If
\[
\prod_{x,y}g_{E_{xy}}>0,
\]
then
\[
h_{\mathrm L}(\vec g)>\vec g\cdot \vec p.
\]
\end{lemma}

A fixed-marginal local-model construction of this
kind also appears in Ref.~\cite{SunWangYu2026}.
We give the argument here for completeness.

\begin{proof}
First suppose that all four correlator coefficients
are strictly positive. Positivity of the
probabilities $P(+-|xy)$ and $P(-+|xy)$ gives
\[
E_{xy}(\vec p)\le 1-|\alpha_x-\beta_y|.
\]

Construct a local behaviour $\vec q$ from a shared
random variable $U$, uniformly distributed on
$[0,1]$, with deterministic response functions
\[
A_x(U)=
\begin{cases}
+1,&U\le(1+\alpha_x)/2,\\
-1,&U>(1+\alpha_x)/2,
\end{cases}
\]
and
\[
B_y(U)=
\begin{cases}
+1,&U\le(1+\beta_y)/2,\\
-1,&U>(1+\beta_y)/2.
\end{cases}
\]
This behaviour has the same marginals as $\vec p$.
Writing $p_A=(1+\alpha_x)/2$ and
$p_B=(1+\beta_y)/2$ for the two thresholds, the
probability that the two outcomes agree is
$1-|p_A-p_B|$, and therefore
\[
E_{xy}(\vec q)=1-2|p_A-p_B|=1-|\alpha_x-\beta_y|.
\]
Consequently,
\[
\vec g\cdot \vec q-\vec g\cdot \vec p
=
\sum_{x,y}g_{E_{xy}}
\bigl[1-|\alpha_x-\beta_y|-E_{xy}(\vec p)\bigr]
\ge0.
\]
Since every $g_{E_{xy}}>0$, equality requires
equality of all four correlators. Together with
equality of the marginals, this would imply
$\vec p=\vec q$, contradicting nonlocality of
$\vec p$. Thus
\[
h_{\mathrm L}(\vec g)\ge \vec g\cdot \vec q>\vec g\cdot \vec p.
\]

Now suppose only that $\prod_{x,y}g_{E_{xy}}>0$.
Writing $s_{xy}=\operatorname{sign}(g_{E_{xy}})$,
choose local outcome flips
\begin{align*}
\epsilon_0&=1,&
\zeta_0&=s_{00},\\
\zeta_1&=s_{01},&
\epsilon_1&=s_{10}s_{00}.
\end{align*}
The first three pairs satisfy
$\epsilon_x\zeta_y s_{xy}=1$ immediately. For the
remaining pair,
\[
\epsilon_1\zeta_1 s_{11}
=s_{10}s_{00}s_{01}s_{11}
=\prod_{x,y}s_{xy}=1,
\]
since the product of the four signs is positive.
Flipping Alice's outcome on input $x$ according to
$\epsilon_x$, and Bob's outcome on input $y$
according to $\zeta_y$, changes the sign of
$g_{E_{xy}}$ and of $E_{xy}$, and likewise changes
the signs of the corresponding marginal
coefficients, while leaving every Bell value
unchanged. After these flips, all four correlator
coefficients are positive, and the preceding
argument applies. Locality and nonlocality are
likewise preserved. This proves the claim.
\end{proof}

\begin{corollary}
\label{cor:candidate-sign}
Let $\vec p$ be nonlocal, and suppose that
$\vec g\cdot \vec p=1$ and
$h_{\mathrm L}(\vec g)\le1$. Then
\[
\prod_{x,y}g_{E_{xy}}\le0.
\]
\end{corollary}

\begin{proof}
A strictly positive product would contradict
Lemma~\ref{lem:local-domination}.
\end{proof}

\label{app:suff-close}

Let $\vec p$ be a nonlocal extreme behaviour.
It admits a planar two-qubit realization with
non-commuting measurements on both sides.
The uniqueness of the overlap $c_0\in(-1,1)$ and
the saturation of the four ellipse constraints were
established above. The necessary inclusion
\[
N_{\mathcal Q}(\vec p)\subseteq\mathcal C(\vec p)
\]
is Eq.~\eqref{eq:B17}.

For the converse, take a nonzero
$\vec g\in\mathcal C(\vec p)$.
Write
\[
\vec g=\sum_{a,x}\mu_{a|x}\vec\gamma_{a|x},
\qquad \mu_{a|x}\ge0,
\]
and put
\[
\lambda=\sum_{a,x}\mu_{a|x}.
\]
Since $\vec g\ne0$, one has $\lambda>0$, and
Lemma~\ref{lem:kernel} gives
$\vec g\cdot \vec p=\lambda$.
The normalized expression
$\widehat{\vec g}=\vec g/\lambda$ satisfies
\[
\widehat{\vec g}\cdot \vec p=1,\qquad
h_{\mathrm L}(\widehat{\vec g})\le1,
\]
and retains the overlap-stationarity condition.

Corollary~\ref{cor:candidate-sign} implies
\[
\prod_{x,y}\widehat g_{E_{xy}}\le0.
\]
Proposition~\ref{prop:first-order} therefore gives
\[
h_{\mathcal Q}(\widehat{\vec g})=1.
\]
By positive homogeneity,
\[
h_{\mathcal Q}(\vec g)
=\lambda h_{\mathcal Q}(\widehat{\vec g})
=\lambda
=\vec g\cdot \vec p.
\]
Thus $\vec g\in N_{\mathcal Q}(\vec p)$.
The zero expression is trivially supporting, so
\[
\mathcal C(\vec p)\subseteq N_{\mathcal Q}(\vec p).
\]
Together with the necessary inclusion, this proves
Theorem~\ref{thm:dualcone}.

\section{Bell values and device-independent randomness certification}
\label{app:own}

This appendix relates randomness certification
to the choice of a Bell expression and an honest
device.
We establish when a randomness-certifying score
must support the set of allowed honest behaviours,
and apply these results to detector loss.

\subsubsection{Guessing probability}

Eve guesses Alice's output for input $x=1$.
Her guessing measurement induces a decomposition
of the observed behaviour into quantum behaviours
$\vec z_e\in\mathcal Q$, with probabilities $t_e$
and guesses $e\in\{-1,+1\}$.
Write
\[
\Pi_e(\vec z)=\frac{1+e\alpha_1(\vec z)}2.
\]
The full-statistics guessing probability is
\[
\Pguess(\vec p)
=
\max
\left\{
\sum_e t_e\Pi_e(\vec z_e):
\sum_e t_e\vec z_e=\vec p
\right\},
\]
where the maximization is over
$t_e\geq0$, $\sum_e t_e=1$, and
$\vec z_e\in\mathcal Q$.
This formulation includes arbitrary quantum side
information: any attack induces such a decomposition,
and conversely any such decomposition can be realized
with a flag held by Eve.

For a Bell expression $\vec g$ and a feasible
threshold $\omega\leq h_{\mathcal Q}(\vec g)$,
$\Pguess(\vec g,\omega)$ is the same maximization
with the full-statistics constraint replaced by
\[
\vec g\cdot\sum_e t_e\vec z_e\geq\omega.
\]
Thus only the average score is retained.
The attack set is always the unrestricted quantum
set $\mathcal Q$, including when the honest
experiment has detector loss.

We say that $\vec g$ is \emph{tight at $\vec p$} if
\[
\Pguess(\vec g,\vec g\cdot\vec p)
=\Pguess(\vec p).
\]
A tight score retains all the information relevant
to this guessing probability, although it need not
determine the full behaviour.

\begin{lemma}[Basic properties]
\label{lem:props}
The following statements hold.
\begin{enumerate}
\item[(a)]
For every feasible threshold,
\[
\Pguess(\vec g,\omega)
=
\sup_{\substack{\vec q\in\mathcal Q\\
                 \vec g\cdot\vec q\geq\omega}}
\Pguess(\vec q).
\]
In particular,
$\Pguess(\vec p)\leq
\Pguess(\vec g,\vec g\cdot\vec p)$.

\item[(b)]
The function $\vec p\mapsto\Pguess(\vec p)$ is
concave on $\mathcal Q$.
The function
$\omega\mapsto\Pguess(\vec g,\omega)$ is concave
and non-increasing on
$(-\infty,h_{\mathcal Q}(\vec g)]$.

\item[(c)]
If $\omega\leq h_{\mathrm L}(\vec g)$, then
$\Pguess(\vec g,\omega)=1$.

\item[(d)]
If $\Pguess(\vec g,\omega^*)<1$, then
$\omega^*>h_{\mathrm L}(\vec g)$ and, for
$\omega^*\leq\omega\leq h_{\mathcal Q}(\vec g)$,
\[
\Pguess(\vec g,\omega)
\leq
\Pguess(\vec g,\omega^*)
-\kappa(\omega-\omega^*),
\]
where
\[
\kappa=
\frac{1-\Pguess(\vec g,\omega^*)}
{\omega^*-h_{\mathrm L}(\vec g)}>0.
\]
\end{enumerate}
\end{lemma}

\begin{proof}
(a) Every attack compatible with the score has
an average behaviour $\vec q\in\mathcal Q$ with
$\vec g\cdot\vec q\geq\omega$.
Conversely, every decomposition of such a
$\vec q$ is compatible with the score.

(b) Mixing two attacks mixes their average
behaviours, scores, and success probabilities.
This proves both concavity statements.
Increasing the threshold reduces the feasible
attack set, proving monotonicity.

(c) A local deterministic behaviour attaining
$h_{\mathrm L}(\vec g)$ is an admissible,
perfectly guessable attack.

(d) Part (c) gives
$\omega^*>h_{\mathrm L}(\vec g)$.
For $\omega>\omega^*$, write
\[
\omega^*=
\lambda h_{\mathrm L}(\vec g)+(1-\lambda)\omega,
\qquad 0<\lambda<1.
\]
Concavity and part (c) imply
\[
\Pguess(\vec g,\omega^*)
\geq
\lambda+(1-\lambda)\Pguess(\vec g,\omega).
\]
Rearranging gives the stated estimate.
The case $\omega=\omega^*$ is immediate.
\end{proof}

\subsubsection{Optimal devices and supporting scores}

Let $K\subseteq\mathcal Q$ be the set of allowed
honest behaviours. Minimizing $\Pguess$ on $K$
is equivalent to maximizing the single-round
min-entropy certified from the full behaviour.

\begin{theorem}[Support at a randomness optimum]
\label{thm:own}
Let $K\subseteq\mathcal Q$ be convex.
Suppose $\vec p^*\in K$ is a local minimizer of
$\Pguess$ on $K$, with $\Pguess(\vec p^*)<1$.
If $\vec g$ is tight at $\vec p^*$, then
\[
\vec g\cdot\vec p\leq\vec g\cdot\vec p^*
\qquad\text{for every }\vec p\in K.
\]
Thus $\vec g$ supports the honest set at $\vec p^*$.
\end{theorem}

\begin{proof}
Set $\omega^*=\vec g\cdot\vec p^*$.
Tightness gives
$\Pguess(\vec g,\omega^*)=\Pguess(\vec p^*)<1$.
If some $\vec p\in K$ had
$\vec g\cdot\vec p>\omega^*$, then convexity would
give
\[
\vec p_t=(1-t)\vec p^*+t\vec p\in K,
\qquad 0<t\leq1.
\]
Its score is strictly larger than $\omega^*$,
so Lemma~\ref{lem:props} implies
\[
\Pguess(\vec p_t)
\leq\Pguess(\vec g,\vec g\cdot\vec p_t)
<\Pguess(\vec g,\omega^*)
=\Pguess(\vec p^*).
\]
Since $\vec p_t\to\vec p^*$ as $t\to0$, this
contradicts local minimality.
\end{proof}

The theorem is a necessary condition for a
full-statistics randomness optimum that admits a
tight score. It does not say that every supporting
score is tight or that every supported behaviour
is randomness-optimal.

For a fixed score, however, maximizing the honest
score is already sufficient to optimize the
single-score certification.

\begin{corollary}[Optimal devices for a fixed score]
\label{cor:single}
Let $K\subseteq\mathcal Q$, and suppose that
$\vec p^*\in K$ maximizes $\vec g\cdot\vec p$
over $K$. Then
\[
\Pguess(\vec g,\vec g\cdot\vec p)
\geq
\Pguess(\vec g,\vec g\cdot\vec p^*)
\qquad\text{for every }\vec p\in K.
\]
In particular, each witness certified in
Theorem~\ref{thm:cones} is optimally implemented,
among honest devices in $\mathcal Q_\eta$, by
its corresponding recorded behaviour.
\end{corollary}

\begin{proof}
The conclusion follows from
$\vec g\cdot\vec p\leq\vec g\cdot\vec p^*$ and
the monotonicity in Lemma~\ref{lem:props}(b).
\end{proof}

This is optimality over honest devices for a
\emph{fixed} score. It is distinct from joint
optimality over devices and scores.

\subsubsection{Optimization under detector loss}

For independent equal-efficiency detectors, with
every missed detection recorded as $+1$, the
honest loss map is
\[
L_\eta(\vec q)=\Lambda_\eta\vec q+\vec k_\eta,
\qquad
\mathcal Q_\eta=L_\eta(\mathcal Q),
\]
as defined in
Eqs.~\eqref{eq:lossmarg}--\eqref{eq:lossLambda}.
For $0<\eta\leq1$, this is an invertible affine
map; at $\eta=1$ it is the identity.
Moreover,
\[
L_r\circ L_\eta=L_{r\eta}.
\]
Since $L_r$ is a local stochastic operation,
$L_r(\mathcal Q)\subseteq\mathcal Q$.
Consequently,
$\mathcal Q_\eta\subseteq\mathcal Q_{\eta'}$
whenever $0<\eta\leq\eta'\leq1$.
Operationally, a more efficient experiment can
emulate a less efficient one by independently
discarding clicks.

\begin{proposition}[Supporting scores under loss]
\label{prop:reduce}
Let $0<\eta\leq1$.
\begin{enumerate}
\item[(i)]
If $\inf_{\mathcal Q_\eta}\Pguess<1$ is attained,
then it is attained at
$L_\eta(\vec q^*)$ for some nonlocal extreme
$\vec q^*\in\mathcal Q$, which admits a
two-qubit realization.

\item[(ii)]
A score $\vec g$ supports $\mathcal Q_\eta$
at $L_\eta(\vec q)$ if and only if
$\Lambda_\eta^{\mathsf T}\vec g$
supports $\mathcal Q$ at $\vec q$.

\item[(iii)]
At a randomness-optimal behaviour
$L_\eta(\vec q^*)$ as in (i), every tight score
$\vec g$ satisfies
\[
\Lambda_\eta^{\mathsf T}\vec g
\in N_{\mathcal Q}(\vec q^*)
=\mathcal C(\vec q^*).
\]
\end{enumerate}
\end{proposition}

\begin{proof}
(i) The affine bijection $L_\eta$ maps extreme
points of $\mathcal Q$ to extreme points of
$\mathcal Q_\eta$.
Every point of this compact finite-dimensional
convex set is a finite convex combination of
extreme points~\cite[Cor.~18.5.1]{Rockafellar}.
If a minimizing behaviour has decomposition
$\vec p^*=\sum_i\lambda_i\vec e_i$, concavity gives
\[
\Pguess(\vec p^*)
\geq\sum_i\lambda_i\Pguess(\vec e_i)
\geq\inf_{\mathcal Q_\eta}\Pguess.
\]
Both inequalities are equalities, so every
positive-weight component is also minimizing.
A local ideal behaviour remains local after
loss, and every local behaviour has guessing
probability one. A minimizing extreme component
with value below one therefore has a nonlocal
extreme ideal preimage.

(ii) For every $\vec z\in\mathcal Q$,
\[
\vec g\cdot L_\eta(\vec z)
=
(\Lambda_\eta^{\mathsf T}\vec g)\cdot\vec z
+\vec g\cdot\vec k_\eta.
\]
The last term is independent of $\vec z$.

(iii) Apply Theorem~\ref{thm:own} to
$K=\mathcal Q_\eta$, then use (ii) and
Theorem~\ref{thm:dualcone}.
\end{proof}

The relevant ideal normal is therefore the
pulled-back expression
$\Lambda_\eta^{\mathsf T}\vec g$, not generally
the recorded-data expression $\vec g$.
Support over the honest set $\mathcal Q_\eta$
must not be confused with support over
$\mathcal Q$.

The preceding necessary condition assumes an
exactly tight score. The following formulation
concerns optimal values and does not require
attainment by a finite score.

Define the full-statistics guessing bias
\[
\Gamma(\vec p)=2\Pguess(\vec p)-1,
\]
and let $\vec m$ be the coordinate vector satisfying
$\vec m\cdot\vec p=\alpha_1$.
For $\vec g\in\mathbb R^8$, set
\[
U(\vec g)
=
\max_{a=\pm1}h_{\mathcal Q}(\vec g+a\vec m).
\]
The standard guessing-probability
dual~\cite{NietoSilleras2014} is
\[
\Gamma(\vec p)
=
\inf_{\vec g}
\bigl[U(\vec g)-\vec g\cdot\vec p\bigr].
\]
Indeed, $\Gamma$ is the concave envelope of
$\vec q\mapsto|\vec m\cdot\vec q|$ on
$\mathcal Q$, and an affine function
$u-\vec g\cdot\vec q$ majorizes this absolute
bias throughout $\mathcal Q$ precisely when
$u\geq U(\vec g)$.
The dual infimum need not be attained at a
finite $\vec g$.

\begin{corollary}[Joint optimization]
\label{cor:value}
Let $0<\eta\leq1$ and $K=\mathcal Q_\eta$.
Then
\begin{equation}
\label{eq:valuecomplete}
\begin{split}
\inf_{\vec p\in K}\Gamma(\vec p)
=
\inf_{\vec g}\Bigl[
&\max_{a=\pm1}h_{\mathcal Q}(\vec g+a\vec m)\\
&-h_{\mathcal Q}(\Lambda_\eta^{\mathsf T}\vec g)
-\vec g\cdot\vec k_\eta
\Bigr].
\end{split}
\end{equation}
If this value is strictly below one, the infimum
is unchanged when restricted to expressions
satisfying
\[
\Lambda_\eta^{\mathsf T}\vec g
\in\mathcal C(\vec q)
\]
for some nonlocal extreme maximizer $\vec q$
of $\Lambda_\eta^{\mathsf T}\vec g$ on
$\mathcal Q$.
\end{corollary}

\begin{proof}
Using the full-statistics dual and interchanging
the two infima gives
\[
\begin{aligned}
\inf_{\vec p\in K}\Gamma(\vec p)
&=
\inf_{\vec p\in K}\inf_{\vec g}
[U(\vec g)-\vec g\cdot\vec p]\\
&=
\inf_{\vec g}[U(\vec g)-h_K(\vec g)].
\end{aligned}
\]
The support function of the affine image is
\[
h_K(\vec g)
=
h_{\mathcal Q}(\Lambda_\eta^{\mathsf T}\vec g)
+\vec g\cdot\vec k_\eta,
\]
proving Eq.~\eqref{eq:valuecomplete}.

At every local deterministic behaviour $\vec d$,
$|\vec m\cdot\vec d|=1$. Hence
\[
U(\vec g)\geq h_{\mathrm L}(\vec g)+1.
\]
If $U(\vec g)-h_K(\vec g)<1$, it follows that
$h_K(\vec g)>h_{\mathrm L}(\vec g)$.
No local ideal behaviour can then maximize
$\Lambda_\eta^{\mathsf T}\vec g$, because its
recorded behaviour remains local.
Choose an extreme maximizer $\vec q$.
It is nonlocal, and Theorem~\ref{thm:dualcone}
gives
$\Lambda_\eta^{\mathsf T}\vec g\in\mathcal C(\vec q)$.
When the infimum is below one, expressions with
objective value below one suffice to approach it.
\end{proof}

Theorem~\ref{thm:dualmain} evaluates the quantum
support functions in this formula, while
Theorem~\ref{thm:dualcone} characterizes the
allowed pulled-back normals.
This is a reduction of the optimal-value problem,
not a claim that the remaining optimization is
convex or that the reported numerical devices
attain its global optimum.

\subsubsection{Tight scores}

We finally distinguish the normal cone of the
honest set from the normal cone of the unrestricted
quantum set at the same recorded behaviour.

\begin{proposition}[Normal cones after loss]
\label{prop:losscone}
Let $\vec q\in\mathcal Q$ have strictly
nondeterministic marginals, and let
$0<\eta_A,\eta_B<1$.
Suppose missed detections are recorded as $+1$,
and write
$\vec p=L_{\eta_A,\eta_B}(\vec q)$.
Then
\[
N_{\mathcal Q}(\vec p)
=
\operatorname{cone}
\{\vec n_{xy}:P_{\vec q}(--|xy)=0\},
\]
where
\[
\vec n_{xy}\cdot\vec z
=
\alpha_x(\vec z)+\beta_y(\vec z)-E_{xy}(\vec z).
\]
Consequently, $\vec p$ is interior to
$\mathcal Q$ if and only if all four
$P_{\vec q}(--|xy)$ are positive.
In that case there exists a finite score
$\vec g_{\vec p}$ that is exactly tight at
$\vec p$.

For every nonlocal $\vec q$ in this scenario,
the marginal assumption holds automatically.
\end{proposition}

\begin{proof}
Set $a=\eta_A\eta_B\in(0,1)$ and decompose
\[
\vec p=a\vec q+(1-a)\vec r,
\]
where $\vec r$ is the normalized mixture of
the three click patterns in which at least
one detector fails.
This behaviour is local and satisfies
$P_{\vec r}(--|xy)=0$ for all $x,y$.

The positive single-loss and double-loss weights,
together with the strictly nondeterministic
marginals of $\vec q$, imply that the other
three probabilities at every input pair are
strictly positive.
Every CHSH inequality is also strict at
$\vec r$: in a single-loss component its
absolute value is twice the absolute value
of a nondeterministic marginal, hence below
two, whereas the double-loss component has
absolute value two.
The local polytope has only positivity and
CHSH facets, so
\[
N_{\mathrm L}(\vec r)
=
\operatorname{cone}\{\vec n_{xy}:x,y\in\{0,1\}\}.
\]
Since $\mathrm L\subseteq\mathcal Q$,
$N_{\mathcal Q}(\vec r)\subseteq
N_{\mathrm L}(\vec r)$.
Conversely, every $\vec n_{xy}$ is a quantum
supporting normal at $\vec r$, by positivity:
\[
\vec n_{xy}\cdot\vec z
=
1-4P_{\vec z}(--|xy)\leq1.
\]
Thus the two normal cones at $\vec r$ coincide.

For a convex set, the normal cone at a strict
convex combination is the intersection of the
normal cones at its endpoints. Therefore
\[
N_{\mathcal Q}(\vec p)
=
N_{\mathcal Q}(\vec q)\cap
N_{\mathcal Q}(\vec r).
\]
For
$\vec g=\sum_{xy}\lambda_{xy}\vec n_{xy}$,
$\lambda_{xy}\geq0$, all positivity bounds
are simultaneously attained at $\vec r$, so
\[
h_{\mathcal Q}(\vec g)=\sum_{xy}\lambda_{xy}.
\]
Consequently,
\[
h_{\mathcal Q}(\vec g)-\vec g\cdot\vec q
=
4\sum_{xy}\lambda_{xy}P_{\vec q}(--|xy).
\]
This vanishes exactly when every positive
$\lambda_{xy}$ corresponds to a zero ideal
probability, proving the cone identity.
Since $\mathcal Q$ is full-dimensional,
$\vec p$ is interior exactly when its normal
cone is $\{0\}$.

Suppose now that $\vec p$ is interior.
There exists $d>0$ such that
$\vec p+dB_2\subseteq\mathcal Q$, where $B_2$
is the Euclidean unit ball.
Convexity of the support function gives
\[
U(\vec g)
=\max_{a=\pm1}h_{\mathcal Q}(\vec g+a\vec m)
\geq h_{\mathcal Q}(\vec g).
\]
Hence
\[
U(\vec g)-\vec g\cdot\vec p
\geq
h_{\mathcal Q}(\vec g)-\vec g\cdot\vec p
\geq d\|\vec g\|_2.
\]
The dual objective is continuous and coercive,
so it attains its minimum at a finite
$\vec g_{\vec p}$.

The resulting affine certificate gives
\[
\Gamma(\vec z)
\leq U(\vec g_{\vec p})
-\vec g_{\vec p}\cdot\vec z
\leq U(\vec g_{\vec p})
-\vec g_{\vec p}\cdot\vec p
=\Gamma(\vec p)
\]
whenever
$\vec g_{\vec p}\cdot\vec z
\geq\vec g_{\vec p}\cdot\vec p$.
Thus the single-score guessing probability
is at most $\Pguess(\vec p)$.
The reverse inequality follows from
Lemma~\ref{lem:props}(a), proving exact tightness.
\end{proof}

\section{Certificates under detection loss}
\label{app:lossy}

This appendix separates the numerical search for
honest devices and Bell scores from the rigorous
verification of their properties.
The search identifies candidate strategies,
supporting expressions, and attacks.
The published conclusions are then established
using exact rational arithmetic and
outward-rounded interval arithmetic.

There are three distinct certificates.
First, exact linear calculations determine each
candidate supporting polygon, and independent
quantum-bound certificates verify its vertices.
Second, NPA dual certificates bound Eve's guessing
probability from above, while explicit compatible
attacks bound it from below.
Third, certified CHSH bounds and loss emulation
establish the comparison throughout the efficiency
interval.

The independent vertex certificates provide an
additional verification of the numerical examples.
They are not an extra hypothesis of the general
normal-cone characterization in
Theorem~\ref{thm:dualcone}.
The certificate inputs, verification procedures,
and reported outputs are identified in the
ancillary archive~\cite{anc}.

\subsubsection{CHSH bounds}

Write
\[
\mathrm{CHSH}=E_{00}+E_{01}+E_{10}-E_{11},
\]
and define
\[
c_Q(\alpha,\alpha)
=
\max_{\vec q\in\mathcal Q}
\bigl[
\mathrm{CHSH}(\vec q)
+\alpha\alpha_0(\vec q)
+\alpha\beta_0(\vec q)
\bigr].
\]
Ref.~\cite[Theorem~1]{Gigena2025} determines this
doubly tilted quantum bound analytically.
Explicitly, $c_Q(\alpha,\alpha)$ is the largest real
root of the degree-4 polynomial identified in that
theorem.

\begin{theorem}[CHSH under detector loss]
\label{thm:smax}
Let $\eta\in(2/3,1]$ and
$\alpha=2(1-\eta)/\eta$.
With every missed detection recorded as $+1$,
the largest recorded value of the canonical
CHSH expression is
\begin{equation}
\label{eq:smax}
S_{\max}(\eta)
=
2(1-\eta)^2+\eta^2c_Q(\alpha,\alpha).
\end{equation}
This value is attained by a two-qubit strategy.
The largest min-entropy certified from this
canonical score is therefore
\[
H_{\min}^{\mathrm{CHSH}}(\eta)
=
-\log_2\!\left[
\frac{1+\sqrt{2-S_{\max}(\eta)^2/4}}2
\right].
\]
\end{theorem}

\begin{proof}
Substituting the loss relations gives
\[
\begin{aligned}
\mathrm{CHSH}(L_\eta(\vec q))
={}&2(1-\eta)^2\\
&+\eta^2\bigl[
\mathrm{CHSH}(\vec q)
+\alpha\alpha_0(\vec q)
+\alpha\beta_0(\vec q)
\bigr].
\end{aligned}
\]
Maximizing over $\mathcal Q$ gives
Eq.~\eqref{eq:smax}, with attainability supplied
by Ref.~\cite{Gigena2025}.
The exact CHSH guessing-probability formula
of Eq.~\eqref{eq:hminCHSH} is decreasing in
the score before conversion to min-entropy.
Thus the largest observed Bell value gives the
largest single-score min-entropy.
\end{proof}

The benchmark allows every honest quantum device,
every CHSH sign pattern, and every fixed,
input-dependent assignment of a missed detection
to an outcome.
Explicitly, a CHSH variant is
\[
S_s=\sum_{x,y}s_{xy}E'_{xy},
\qquad
s_{xy}\in\{-1,+1\},
\qquad
\prod_{x,y}s_{xy}=-1.
\]
Alice's missed detection on input $x$ is assigned
a fixed outcome $a_x\in\{-1,+1\}$, and Bob's
on input $y$ a fixed outcome $b_y\in\{-1,+1\}$.
Eve is constrained only by the observed Bell value,
and her attack set remains $\mathcal Q$.

Relabelling the outputs by $a_x$ and $b_y$
makes all no-click assignments equal to $+1$.
The transformed CHSH signs are
$s'_{xy}=s_{xy}a_xb_y$, whose product is still
$-1$.
Input permutations then reduce the possibilities
to two classes: one negative sign or three
negative signs.

\begin{proposition}[CHSH variants]
\label{prop:variants}
For every CHSH variant, the min-entropy certified
from a score threshold $S$ is
\[
f(S)
=
-\log_2\!\left[
\frac{1+\sqrt{2-S^2/4}}2
\right]
\]
for $2\leq S\leq2\sqrt2$, and is zero for
$S\leq2$.

After making all no-click assignments equal to
$+1$, the one-negative-sign class has maximum
$S_{\max}(\eta)$ of Theorem~\ref{thm:smax}.
At the six reference efficiencies and every
efficiency in the continuum certificate grid,
the three-negative-sign class has a certified
strictly smaller maximum.
The minimum reported certified gap is
$3.19\times10^{-3}$.
\end{proposition}

\begin{proof}
Input and output relabellings preserve the
quantum attack set and guessing probability,
so the exact CHSH formula applies to every
variant.
For $S\leq2$, a local deterministic attack attaining
score two is feasible and perfectly guessable.

The reduction of no-click assignments and sign
patterns was described above.
For the three-negative-sign class, an exact
rational NPA dual certificate supplies an
upper bound on the largest observed Bell value.
At every certified efficiency this upper bound
lies below a certified lower bound on the
canonical maximum.
The archive records the certificates and the
resulting gaps~\cite{anc}.
\end{proof}

\subsubsection{Supporting polygons}

For each reference efficiency, the numerical
search uses a state
\[
|\psi_\vartheta\rangle
=
\cos\vartheta|00\rangle+\sin\vartheta|11\rangle
\]
and coplanar projective measurements.
It searches jointly over the device and over
scores whose pullbacks belong to the candidate
supporting family.

To obtain exact input data, the pairs
$(\cos2\vartheta,\sin2\vartheta)$ and the
measurement direction cosines and sines are
replaced by rational points on the unit circle.
The resulting ideal behaviour $\vec q_\eta$,
Bob's overlap, the four tangent expressions,
and the stationarity coefficients are therefore
exactly rational.
The entanglement parameters found numerically
range from approximately $0.092\pi$ at
$\eta=0.75$ to $0.248\pi$ at $\eta=0.98$.

At each stored device, all four fixed-overlap
constraints are saturated.
The nonnegative tangent weights, normalization,
measurement stationarity, and sixteen local
comparisons define a rational candidate polygon.
After determining its affine dimension, the
vertices are enumerated in exact arithmetic
by intersecting the boundary constraints and
retaining the feasible solutions.

For an independent check of global support,
each vertex $\vec V$ is supplied with a
two-component function $\vec\delta_{\vec V}(c)$
such that
\[
\Phi_c(\vec V,\vec\delta_{\vec V}(c))\leq1
\qquad\text{for every }c\in[-1,1].
\]
The target normalization is
$\vec V\cdot\vec q_\eta=1$.
The certificate functions have the form
\[
\begin{aligned}
\vec\delta_{\vec V}(c)
={}&\vec d_0+\vec d_1(c-c_0)\\
&+(c-c_0)^2
\left[
\sum_j\vec R_jT_j(c)+\vec a+\vec b\,c
\right],
\end{aligned}
\]
where $T_j$ are Chebyshev polynomials and all
stored vector coefficients are rational.

To state what is verified, define
\[
\ell_{x,a}(c)
=
aV_{\alpha_x}
+
\left\|
\frac{\vec V_\beta}{2}
+(-1)^x\vec\delta_{\vec V}(c)
+a\vec V_{E_x}
\right\|_c.
\]
For $\boldsymbol a=(a_0,a_1)\in\{-1,+1\}^2$,
let
\[
F_{\boldsymbol a}(c)
=
1-\ell_{0,a_0}(c)-\ell_{1,a_1}(c).
\]
Since
\[
\Phi_c(\vec V,\vec\delta_{\vec V}(c))
=
\max_{a_0,a_1}
\bigl[\ell_{0,a_0}(c)+\ell_{1,a_1}(c)\bigr],
\]
the required bound is equivalent to
nonnegativity of these four margins.

The stored coefficients satisfy the contact
conditions
\[
F_{\boldsymbol a}(c_0)=0,
\qquad
F'_{\boldsymbol a}(c_0)=0,
\]
which are checked exactly, together with the
endpoint inequalities.
The remaining intervals are verified using
160-bit ball arithmetic~\cite{Arb}, with
subdivision and direct enclosures, mean-value
bounds, positive second-derivative bounds near
$c_0$, or monotonicity towards an endpoint.
Thus the verification covers every real overlap,
not only a finite sample.
Theorem~\ref{thm:dualmain} then gives
$h_{\mathcal Q}(\vec V)\leq1$.

\begin{theorem}[Supporting polygons]
\label{thm:cones}
For each of the six reference devices, the
normal cone at the stored ideal behaviour is
\[
N_{\mathcal Q}(\vec q_\eta)
=
\mathcal C(\vec q_\eta).
\]
Its section normalized by
$\vec g\cdot\vec q_\eta=1$ is the rational
polygon with $n_\eta$ vertices reported in
Table~\ref{tab:lossy}.
Consequently,
\[
N_{\mathcal Q_\eta}(L_\eta(\vec q_\eta))
=
\Lambda_\eta^{-\mathsf T}
\mathcal C(\vec q_\eta).
\]
\end{theorem}

\begin{proof}
Saturation of the four fixed-overlap constraints
and strict feasibility of their defining convex
system imply that its normal cone is generated
by the four tangent expressions.
Any expression supporting $\mathcal Q$ at
$\vec q_\eta$ must also satisfy measurement
stationarity and the sixteen local comparisons.
Hence it belongs to $\mathcal C(\vec q_\eta)$.

Exact vertex enumeration determines the full
normalized candidate polygon.
For every listed vertex $\vec V$, the independent
certificate above proves
$h_{\mathcal Q}(\vec V)\leq1$.
Since $\vec V\cdot\vec q_\eta=1$, equality holds.
Convex combinations of these normalized
expressions also support at $\vec q_\eta$,
as do their nonnegative rescalings.
This proves the reverse inclusion.

The identity for the honest lossy set follows
from Proposition~\ref{prop:reduce}(ii).
\end{proof}

\subsubsection{Randomness bounds}

For each device we select
\[
\vec g_\eta
=
\Lambda_\eta^{-\mathsf T}
\sum_i\lambda_i\vec V_i,
\qquad
\lambda_i\geq0,
\qquad
\sum_i\lambda_i=1,
\]
with rational weights.
Its honest score
\[
\omega_\eta
=
\vec g_\eta\cdot L_\eta(\vec q_\eta)
\]
is therefore exactly rational.

Let $\vec m$ satisfy
$\vec m\cdot\vec z=\alpha_1(\vec z)$.
For every $\lambda\geq0$,
\begin{equation}
\label{eq:lossy-guess-dual}
\Pguess(\vec g,\omega)
\leq
\frac12+
\max_{e=\pm1}
h_{\mathcal Q}\!\left(
\lambda\vec g+\frac e2\vec m
\right)
-\lambda\omega.
\end{equation}
Indeed, for any compatible attack,
adding
$\lambda(\vec g\cdot\sum_e t_e\vec z_e-\omega)$
to its success probability can only increase
that probability.
Each resulting conditional Bell value is at
most the corresponding quantum support function.

The support functions in
Eq.~\eqref{eq:lossy-guess-dual} are bounded by
exact rational NPA dual certificates at the
stored relaxation levels~\cite{NPA2007,NPA2008}.
The verification checks the dual coefficient
identities and objective values in exact
arithmetic, and positive semidefiniteness by
an exact $LDL^{\mathsf T}$ factorization.
This gives an upper bound $P_U$ on Eve's
guessing probability, and hence the lower
entropy bound $-\log_2P_U$.

For the opposite direction, an explicit flagged
mixture of two-qubit strategies gives a quantum
attack.
Its normalization, physical validity, and score
feasibility are verified, in particular
\[
\sum_e t_e\vec g_\eta\cdot\vec z_e
\geq\omega_\eta.
\]
An outward-rounded lower bound $P_L$ on its
success probability gives an upper bound on
the certifiable entropy.
Thus the reported interval is
\[
-\log_2P_U
\leq
-\log_2\Pguess(\vec g_\eta,\omega_\eta)
\leq
-\log_2P_L.
\]

\begin{theorem}[Randomness advantage; computer-assisted]\label{thm:rates}
For each $\eta$ in Table~\ref{tab:lossy}, the expression $\vec g_\eta=\Lambda_\eta^{-\top}\sum_i\lambda_iV_i$, with rational convex weights over the vertices of Theorem~\ref{thm:cones}, is maximized over $\Q_\eta$ by $\vec p_\eta=L_\eta \vec q_\eta$ and certifies the listed $H_{\min}(A_1|X{=}1,E)$ against every quantum device whose average score is at least that of $\vec p_\eta$. An explicit compatible attack supplies the reported upper bound on the certifiable min-entropy. The CHSH benchmark with the same detectors certifies at most the listed certified CHSH benchmark upper bounds. Those table entries are obtained by applying the guessing-probability formula to a certified NPA upper bound on the recorded canonical CHSH score. The exact canonical maximum under equal-efficiency loss is the analytic value in Theorem~\ref{thm:smax}; it is used, together with NPA certificates for the remaining sign class, in Proposition~\ref{prop:variants}.
\end{theorem}
\begin{proof}
Theorem~\ref{thm:cones} and the choice of
$\vec g_\eta$ imply that
$L_\eta(\vec q_\eta)$ maximizes its score over
$\mathcal Q_\eta$.
The exact NPA dual certificates and the
explicit feasible attacks give the two-sided
entropy intervals as described above.

For the CHSH benchmark, the reported upper
bound is obtained by applying
Eq.~\eqref{eq:hminCHSH} to a certified upper
bound on the canonical observed Bell value.
Proposition~\ref{prop:variants} establishes
that the other sign class cannot exceed the
canonical maximum at these efficiencies.
The resulting entropy upper bounds are
strictly below the selected-score lower bounds.

Finally, Corollary~\ref{cor:single} shows that
each recorded behaviour is optimal among
honest devices for its fixed published score.
No joint optimality over scores is asserted.
\end{proof}

\subsubsection{Intermediate efficiencies}

\begin{table*}[t]
\caption{The six certified reference points. Here $n_\eta$ is the number of vertices in the normalized supporting polygon, $c_0$ is Bob's measurement overlap at the explicit ideal strategy. The last two columns give a two-sided interval
for the selected expression's certifiable
single-round min-entropy and an upper bound
for every CHSH-score protocol, respectively.}
\label{tab:lossy}
\begin{center}
\footnotesize
\setlength{\tabcolsep}{6pt}
\begin{tabular}{ccccc}
\toprule
$\eta$ & $n_\eta$ & $c_0$ & \shortstack{Selected\\expression} & \shortstack{CHSH\\upper bound}\\ \midrule
0.75 & 5 & 0.4578 & $[0.009845,\,0.009847]$ & 0.009013\\
0.80 & 5 & 0.2826 & $[0.037859,\,0.037864]$ & 0.033445\\
0.85 & 7 & 0.1735 & $[0.095104,\,0.095105]$ & 0.081930\\
0.90 & 7 & 0.0935 & $[0.198626,\,0.198627]$ & 0.168313\\
0.95 & 6 & 0.1089 & $[0.385565,\,0.385566]$ & 0.330204\\
0.98 & 8 & 0.3556 & $[0.597367,\,0.597368]$ & 0.525646\\
\bottomrule
\end{tabular}
\end{center}
\end{table*}

Table~\ref{tab:lossyexprs} lists a positive rescaling of
each selected expression $\vec g_\eta$, chosen so that
one correlator coefficient has magnitude one, together
with the correspondingly rescaled observed Bell value.

\begin{table*}[t]
\caption{Selected Bell expressions for the six
reference efficiencies of Table~\ref{tab:lossybrief}.
The coefficient vector is written in the correlator
convention of Eq.~\eqref{eq:gvec}, with outcomes in
$\{\pm1\}$, and the Bell value is
$\mathcal B_{\vec g}(\vec p)=\vec g\cdot\vec p$.
One correlator coefficient is normalized to $\pm1$.
The coefficients and the observed Bell values
$\omega=\vec g\cdot L_\eta(\vec q_\eta)$ are rounded
to eight decimal places; the exact rational values
are the fields \texttt{g} and \texttt{omega} in the
files
\texttt{lossycone/data/t4/}$\langle\eta\rangle$\texttt{.json}
of the ancillary archive~\cite{anc}, with $\eta$
written to three decimal places.}
\label{tab:lossyexprs}
\begin{center}
\footnotesize
\setlength{\tabcolsep}{3.5pt}
\begin{tabular}{ccrrrrrrrr}
\toprule
$\eta$ & $\omega$ & $g_{\alpha_0}$ & $g_{\alpha_1}$ & $g_{\beta_0}$ & $g_{\beta_1}$ & $g_{E_{00}}$ & $g_{E_{01}}$ & $g_{E_{10}}$ & $g_{E_{11}}$\\
\midrule
$0.75$ & $1.88867480$ & $0.01068092$ & $0.14769774$ & $-0.01103777$ & $0.14805473$ & $0.86947105$ & $0.85266685$ & $0.85230245$ & $-1$\\
$0.80$ & $1.90364589$ & $0.01971622$ & $0.20020662$ & $-0.02215746$ & $0.20264756$ & $0.84280748$ & $0.80223478$ & $0.79979380$ & $-1$\\
$0.85$ & $1.87698887$ & $0.02369224$ & $0.31144801$ & $-0.03186308$ & $0.31961907$ & $0.76510273$ & $0.69672721$ & $0.68855210$ & $-1$\\
$0.90$ & $1.47968128$ & $0.00979524$ & $0.67631741$ & $-0.02122317$ & $0.68774541$ & $0.39627149$ & $0.33511062$ & $0.32368263$ & $-1$\\
$0.95$ & $1.80606459$ & $0.00062703$ & $0.55910542$ & $-0.02868180$ & $0.58716022$ & $0.62114817$ & $0.47907760$ & $0.44089459$ & $-1$\\
$0.98$ & $2.17525924$ & $0.52400533$ & $0.07000682$ & $0.27752728$ & $0.34766824$ & $-1$ & $-0.84356646$ & $0.59706389$ & $-0.66707069$\\
\bottomrule
\end{tabular}
\end{center}
\end{table*}

\begin{theorem}[Efficiency interval; computer-assisted]\label{thm:cover}
Let $\eta_1<\dots<\eta_{112}$ run from $0.750$ to $0.980$ (spacing $0.001$--$0.003$, together with $0.90$ and $0.95$). At every $\eta_i$ the conclusions of Theorems~\ref{thm:cones}--\ref{thm:rates} hold for a stored rational device (656 polygon vertices in total), with the CHSH side bounded by certified CHSH benchmark upper bounds. Consequently, for every $\eta\in[0.75,0.98]$ some two-qubit protocol at efficiency $\eta$ certifies more min-entropy per round than every CHSH-score protocol at efficiency $\eta$, by at least $5.3\%$ on $[0.75,0.80]$, $8.5\%$ on $[0.80,0.85]$, $10.7\%$ on $[0.85,0.90]$, $12.2\%$ on $[0.90,0.95]$ and $8.0\%$ on $[0.95,0.98]$; at the grid points the gain is approximately $8.8\%$--$18.0\%$.
\end{theorem}
\begin{proof}
Let $H_i$ be the certified lower entropy bound
for the selected construction at $\eta_i$,
and let $U_i$ be the certified upper bound
for every CHSH-score protocol at $\eta_i$.

Suppose $\eta\in[\eta_i,\eta_{i+1}]$.
On each side, independently retaining each
actual click with probability $\eta_i/\eta$
reproduces the honest recorded behaviour at
efficiency $\eta_i$.
Discarded clicks are treated as missed detections,
and every trial is retained.
The selected protocol therefore certifies
at least $H_i$ at efficiency $\eta$.

Conversely, a laboratory at efficiency
$\eta_{i+1}$ can emulate every honest CHSH
experiment at efficiency $\eta$ by independently
discarding clicks.
This holds for each fixed no-click assignment,
using that same assignment for discarded clicks.
Thus the optimum over all devices, CHSH sign
patterns, and no-click assignments is
non-decreasing with efficiency.
Every CHSH protocol at $\eta$ consequently
certifies at most $U_{i+1}$.

The stored bounds satisfy
$H_i>U_{i+1}$ for every adjacent pair.
Hence the separation holds on every interval,
including its endpoints.
The relative advantage is at least
$H_i/U_{i+1}-1$.
The band minima stated in the theorem are
verified from the certified numerical bounds
by the ancillary script
\texttt{verify\_t4\_rates.py}~\cite{anc}.
\end{proof}

\section{Supporting quadrilateral of the $I_\delta$ family from \cite{WBC2022}}
\label{app:idelta}

This appendix derives the supporting quadrilateral of
the behaviours $\vec p_\delta$ in Section~\ref{sec:idelta}.
Write $s=\sin\delta$ and $c=\cos\delta$, and set
$K=\cos2\delta=2c^2-1$ and $D=2c^3$. Throughout
the interval $0<\delta\leq\pi/6$ one has
$c\in\bigl[\sqrt3/2,1\bigr)$, $s\in\bigl(0,1/2\bigr]$,
and $K\in\bigl[1/2,1\bigr)$.

The unique maximizing behaviour of
$\mathcal I_\delta$ is realized by the measurements of
Eq.~\eqref{eq:Ideltaobs}. The behaviour $\vec p_\delta$
has vanishing marginals, correlators
\(\bigl(0,c,c,-2cs\bigr)\), and Bob overlap
$c_0=-s\in[-1/2,0)$. It is a nonlocal extreme
behaviour: $\mathcal I_\delta$ is uniquely maximized
at this behaviour~\cite{WBC2022}, the measurements are
planar and noncommuting, and $c_0\neq\pm1$.
Theorem~\ref{thm:dualcone} therefore already gives
$N_{\mathcal Q}(\vec p_\delta)=\mathcal C(\vec p_\delta)$.
The remainder of the appendix identifies the
normalized section
$\mathcal P_\delta=\mathcal C(\vec p_\delta)\cap\{\vec g\cdot \vec p_\delta=1\}$.

\subsubsection{Tangent expressions}
The four conditional vectors lie on the ellipse.
With $G_{c_0}^{-1}=(1/c^2)\begin{pmatrix}1&s\\s&1\end{pmatrix}$,
the tangent vectors of Eq.~\eqref{eq:gamma} are
determined by
\begin{equation}
\label{eq:Ideltav}
\begin{aligned}
v_{+|0}&=\bigl(s/c,\,1/c\bigr),&
v_{-|0}&=-v_{+|0},\\
v_{+|1}&=\bigl(K/c,\,-s/c\bigr),&
v_{-|1}&=-v_{+|1}.
\end{aligned}
\end{equation}
A general combination
$\vec g=\sum_{a,x}\mu_{a|x}\vec\gamma_{a|x}$ therefore has
correlator coefficients
\begin{equation}
\label{eq:Ideltacorr}
\begin{aligned}
g_{E_{00}}&=(\mu_{+|0}+\mu_{-|0})\,s/c,\\
g_{E_{01}}&=(\mu_{+|0}+\mu_{-|0})/c,\\
g_{E_{10}}&=(\mu_{+|1}+\mu_{-|1})\,K/c,\\
g_{E_{11}}&=-(\mu_{+|1}+\mu_{-|1})\,s/c,
\end{aligned}
\end{equation}
and Alice's marginal coefficients
$g_{\alpha_0}=-\mu_{+|0}+\mu_{-|0}$,
$g_{\alpha_1}=-\mu_{+|1}+\mu_{-|1}$.
Bob's marginal coefficients are then fixed by the
same weights.

Keep the state, Alice's observables, and $B_0$
fixed, and parametrize Bob's second observable by
\[
B_1(r)=\sqrt{1-r^2}\,\sigma_z+r\sigma_x.
\]
Its overlap with $B_0=\sigma_x$ is $r$, and the
target is recovered at $r=-s$.
The only varying correlators are
\[
E_{01}(r)=\sqrt{1-r^2},
\qquad
E_{11}(r)=-s\sqrt{1-r^2}+cr.
\]
At the target,
\[
\dot E_{01}=\frac{s}{c},
\qquad
\dot E_{11}=\frac{K}{c}.
\]
Consequently, the stationarity condition is
\[
s\,g_{E_{01}}+K\,g_{E_{11}}=0,
\]
which, using Eq.~\eqref{eq:Ideltacorr}, becomes
\begin{equation}
\label{eq:Ideltastat}
\mu_{+|0}+\mu_{-|0}
=
K(\mu_{+|1}+\mu_{-|1}).
\end{equation}

Combined with the normalization
$\vec g\cdot \vec p_\delta=\sum_{a,x}\mu_{a|x}=1$, this fixes
the two sums of weights:
\[
\mu_{+|0}+\mu_{-|0}=\frac{K}{2c^2},\qquad
\mu_{+|1}+\mu_{-|1}=\frac1{2c^2}.
\]
Every candidate therefore has the correlator
coefficients of the normalized expression
$\mathcal I_\delta$,
\begin{equation}
\label{eq:Ideltashared}
\bigl(g_{E_{00}},g_{E_{01}},g_{E_{10}},g_{E_{11}}\bigr)
=\frac1D\bigl(sK,\,K,\,K,\,-s\bigr).
\end{equation}
The remaining freedom is the pair of Alice's
marginal coefficients
$(g_{\alpha_0},g_{\alpha_1})$. Nonnegativity of the four
weights cuts this parameter plane to the rectangle
\begin{equation}
\label{eq:Ideltarect}
\lvert g_{\alpha_0}\rvert\leq A=\frac{K}{2c^2},\qquad
\lvert g_{\alpha_1}\rvert\leq B=\frac1{2c^2}.
\end{equation}

For brevity, write
$u=g_{\alpha_0}$ and $v=g_{\alpha_1}$.
The remaining marginal coefficients are
\begin{equation}
\label{eq:Ideltabob}
g_{\beta_0}=-\frac{su+Kv}{c},
\qquad
g_{\beta_1}=-\frac{u-sv}{c}.
\end{equation}
Thus the entire normalized expression is
determined by the pair $(u,v)$.

\subsubsection{Local bounds}

Consider the local deterministic assignments
\begin{equation}
\label{eq:Ideltalocals}
\begin{aligned}
(a_0,a_1,b_0,b_1)
={}&(1,1,1,1),\;
(1,-1,-1,1),\\
&(-1,1,1,-1),\;
(-1,-1,-1,-1).
\end{aligned}
\end{equation}
Using Eqs.~\eqref{eq:Ideltashared}
and~\eqref{eq:Ideltabob}, their four inequalities
are equivalent to the two strip constraints
\begin{equation}
\label{eq:Ideltastrips}
\begin{aligned}
\bigl|(c-1-s)u+(c-K+s)v\bigr|
&\leq
\frac{(1-c)(c+s)(1+s)}{c^2},\\
\bigl|(c-1+s)u+(K-c+s)v\bigr|
&\leq
\frac{(1-c)(c-s)(1-s)}{c^2}.
\end{aligned}
\end{equation}
For example, the correlator contribution at
$(1,1,1,1)$ is $(K-s^3)/c^3$, and
\[
c^3-K+s^3=(1-c)(c+s)(1+s).
\]
At $(1,-1,-1,1)$ it is $(K+s^3)/c^3$, and
\[
c^3-K-s^3=(1-c)(c-s)(1-s).
\]
The other two inequalities follow by reversing
all four outcomes of a local deterministic assignment.

Both strip widths are strictly positive.
The determinant of their two coefficient
vectors is
\[
-4c^2(1-c)\neq0.
\]
Their intersection is therefore a nondegenerate
centrally symmetric quadrilateral.
Solving the boundary equalities gives exactly
the four vertices in Eq.~\eqref{eq:Ideltaverts},
in cyclic order
$\vec V_1,\vec V_0,\vec V_2,\vec V_3$.

The weights at $\vec V_0$ are
\begin{equation}
\label{eq:Ideltamu0}
\begin{aligned}
\mu_{+|0}&=\frac{K}{4c^3},&
\mu_{-|0}&=\frac{(2c-1)K}{4c^3},\\
\mu_{+|1}&=\frac{c-s}{4c^3},&
\mu_{-|1}&=\frac{c+s}{4c^3},
\end{aligned}
\end{equation}
and those at $\vec V_1$ are
\begin{equation}
\label{eq:Ideltamu1}
\begin{aligned}
\mu_{+|0}&=\frac{(c+s)K}{4c^3},&
\mu_{-|0}&=\frac{(c-s)K}{4c^3},\\
\mu_{+|1}&=\frac{K}{4c^3},&
\mu_{-|1}&=\frac{1+2c-2c^2}{4c^3}.
\end{aligned}
\end{equation}
All are strictly positive on the stated
interval:
$K>0$, $2c-1>0$, $c>s>0$, and
$1+2c-2c^2=1+2c(1-c)>0$.
The weights at the other vertices follow by
interchanging the two outcomes for each input.
Hence the entire quadrilateral lies inside
the weight rectangle, and its vertices lie
strictly inside that rectangle.

It remains to verify that the other twelve
local deterministic inequalities do not cut this
quadrilateral.

It is enough to check $\vec V_0$.
Indeed, exchanging the parties maps $\vec V_0$
to $\vec V_1$, while reversing all marginal
coefficients maps these vertices to
$\vec V_3$ and $\vec V_2$, respectively.
These transformations preserve the local deterministic
comparisons.

Put
\[
h=1-c,\qquad
L=2c+1,\qquad
T=1+c-c^2.
\]
For a local deterministic assignment, the value of
$\vec V_0$, multiplied by $D=2c^3$, is
\[
\begin{aligned}
D\,\vec V_0\cdot\vec d
={}&-sK b_0+hL b_1\\
&+a_0K(-h+s b_0+b_1)\\
&+a_1(s+K b_0-s b_1).
\end{aligned}
\]
We maximize this expression over $a_0,a_1$
for each of Bob's four assignments.

For $(b_0,b_1)=(1,1)$, the coefficients of
$a_0,a_1$ are $K(c+s)$ and $K$, both strictly
positive.
The unique maximizing Alice assignment is
$(a_0,a_1)=(1,1)$, and its value is
\[
hL+K(c+1)=D.
\]
For $(b_0,b_1)=(-1,1)$, the coefficients are
$K(c-s)$ and $-K$.
The unique maximizing assignment is
$(a_0,a_1)=(1,-1)$, again with value $D$.

For $(b_0,b_1)=(1,-1)$, the coefficients are
$K(c+s-2)<0$ and $K+2s>0$.
The maximizing assignment is $(-1,1)$, and
its slack is
\[
D-\max_{a_0,a_1}D\,\vec V_0\cdot\vec d
=
R_-,
\]
where
\[
R_-=4h\bigl[T-s(c+1)\bigr].
\]
For $(b_0,b_1)=(-1,-1)$, the coefficient of
$a_0$ is strictly negative, so $a_0=-1$
maximizes the value.
The slacks for $a_1=-1$ and $a_1=1$ are,
respectively,
\[
R_+=4h\bigl[T+s(c+1)\bigr]
\]
and
\[
R_0=2\bigl(1-2c^2+2c^3-2c^2s\bigr).
\]
Thus the maximal value for this Bob assignment
has slack $\min\{R_+,R_0\}$.

All three quantities $R_-$, $R_+$, and $R_0$
are strictly positive.
For $R_+$ this is immediate.
For $R_-$, both $T$ and $s(c+1)$ are positive,
and
\[
T^2-s^2(c+1)^2=c^2K>0.
\]
Therefore $T>s(c+1)$.

For $R_0$, set
\[
A(c)=1-2c^2+2c^3.
\]
This function is positive and increasing for
$c\in[\sqrt3/2,1)$.
Moreover,
\[
A(c)^2-(2c^2s)^2
=
K\bigl(4c^4-4c^3+2c^2-1\bigr).
\]
The polynomial
$q(c)=4c^4-4c^3+2c^2-1$ satisfies
\[
q'(c)=4c(4c^2-3c+1)>0
\]
for $c>0$, and
\[
q(\sqrt3/2)=\frac{11-6\sqrt3}{4}>0.
\]
Hence $A(c)>2c^2s$, proving $R_0>0$.

It follows that $\vec V_0$ satisfies all
sixteen local deterministic comparisons, with equality
exactly at $(1,1,1,1)$ and $(1,-1,-1,1)$.
The other vertices are feasible by symmetry.
Every local deterministic comparison is affine in
$(u,v)$, so feasibility at all four vertices
proves feasibility throughout the quadrilateral.

Conversely, every candidate expression must
satisfy the four selected comparisons, whose
intersection is this quadrilateral.
The weight constraints and the remaining local
comparisons do not reduce it.
Therefore the normalized candidate section is
exactly the quadrilateral in
Theorem~\ref{thm:idelta}.
By Theorem~\ref{thm:dualcone}, it is the full
normalized supporting family.

\subsubsection{Uniqueness}

The normalized expression $\vec g_I$ associated
with $\mathcal I_\delta$ is the centre of the
quadrilateral.
For every
$\vec g\in\operatorname{ri}\mathcal P_\delta$,
there exist $\vec g'\in\mathcal P_\delta$
and $0<\varepsilon<1$ such that
\[
\vec g
=
\varepsilon\vec g_I+(1-\varepsilon)\vec g'.
\]
Both terms have quantum bound one.
If a quantum behaviour $\vec z$ maximizes
$\vec g$, then
\[
1
=
\varepsilon\vec g_I\cdot\vec z
+(1-\varepsilon)\vec g'\cdot\vec z
\leq1.
\]
Since both weights are positive, equality
requires $\vec g_I\cdot\vec z=1$.
The unique maximizing behaviour of
$\mathcal I_\delta$ is
$\vec p_\delta$~\cite{WBC2022}.
Hence $\vec z=\vec p_\delta$.

Every boundary expression also attains value
one at at least one of the four selected
local deterministic behaviours.
Thus the relative-interior expressions have
a unique quantum maximizer, whereas the
boundary expressions do not.

The algebraic identities in this appendix
can also be checked using the ancillary
script~\cite{anc}.

\section{Comparison of the single Bell score with full statistics}
\label{app:survey}

For a fixed recorded behaviour $\vec p$, retaining
the full statistics can never certify less
randomness than retaining only a Bell score:
\[
\Pguess(\vec p)
\leq
\Pguess(\vec g,\vec g\cdot\vec p).
\]
The reason is that every attack reproducing
$\vec p$ also reproduces its score, whereas the
score alone generally permits additional attacks.
Choosing a Bell expression therefore amounts to
choosing which linear combination of the observed
statistics to retain.

Optimizing over all Bell expressions recovers
the full-statistics guessing probability at the
level of optimal values~\cite{NietoSilleras2014,
Bancal2014}.
This statement does not require that the optimum
be attained by a finite expression.
Nor does it imply that a full-statistics-optimal
expression must belong to the supporting family
of an arbitrarily prescribed honest device.
Appendix~\ref{app:own} gives the precise support
and attainment results relevant to that question.

Table~\ref{tab:fsasym} compares the selected
scores with numerical full-statistics NPA
evaluations at three of the recorded behaviours.
The selected-score values are close to the
full-statistics estimates at these devices.
These numerical comparisons do not establish
exact equality, tightness of the NPA relaxation,
or global optimality over honest devices.
The rigorous two-sided bounds for the selected
scores are those of Table~\ref{tab:lossy}.

\begin{table}[t]
\caption{Single-round min-entropy in bits per
generation round.
Column (a) gives the CHSH-score value at the
CHSH-optimal honest strategy for the stated
detector efficiency.
Columns (b) and (c) concern the same selected
recorded behaviour: (b) uses the published Bell
score, while (c) uses a numerical full-statistics
evaluation at level 3 of the NPA
hierarchy~\cite{NPA2007,NPA2008}.
Thus column (a) generally uses a different
honest device.
The entries are numerical estimates; rigorous
intervals for column (b) and upper bounds for
the CHSH benchmark appear in
Table~\ref{tab:lossy}.}
\label{tab:fsasym}
\centering
\small
\setlength{\tabcolsep}{4pt}
\begin{tabular}{cccc}
\toprule
$\eta$ &
\shortstack{(a) CHSH\\score} &
\shortstack{(b) Selected\\score} &
\shortstack{(c) Full\\statistics}\\
\midrule
0.80 & 0.033444 & 0.037863 & 0.037864\\
0.85 & 0.081930 & 0.095105 & 0.095105\\
0.90 & 0.168312 & 0.198627 & 0.198683\\
\bottomrule
\end{tabular}
\end{table}

\paragraph{Finite statistics.}

The quantities reported here describe
single-round unpredictability.
They are not, by themselves, net randomness
expansion rates or finite-size security bounds.

A single-score analysis and a full-statistics
analysis can use the same recorded events.
Their finite-size performance depends on how
the data are estimated and converted into an
entropy bound, including the input distribution,
confidence region, entropy trade-off function,
and security parameters.
Methods based on several Bell estimators,
entropy accumulation, and quantum probability
estimation address these additional
choices~\cite{SeveralBell,BRC2020,
ZhangKnill2018,ZhangFuKnill2020}.

The improvement established in this work is
therefore an improvement in single-round
certification from one Bell score.
Determining whether it also improves the
extractable randomness in a finite experiment
requires a protocol-level analysis.
The supporting families constructed here provide
a set of Bell expressions over which that further
optimization can be performed.

\section{An asymmetric example illustrating Theorem~\ref{thm:dualcone}}
\label{sec:example}

The family of Section~\ref{sec:idelta} has
vanishing marginals and substantial symmetry.
We now illustrate the construction at a
partially entangled two-qubit behaviour with
nonzero, unequal marginals.

The example is the ideal device used at
$\eta=0.95$ in Section~\ref{sec:lossy}.
Its normalized supporting family is a hexagon.
Besides determining that family, we use it to
answer a concrete question: among all Bell
expressions maximized by this ideal behaviour,
which continue to detect nonlocality at the
lowest detector efficiency?

The state, measurement directions, behaviour,
and supporting expressions are specified by
exact rational data in the archive~\cite{anc}.
The decimal values displayed below are for
readability.

\subsubsection{State and measurements}
The source is the partially entangled state
\begin{equation}
|\psi\rangle=\cos\vartheta|00\rangle+\sin\vartheta|11\rangle,
\end{equation}
with
\begin{equation}
\cos2\vartheta=\frac{73478412}{190096813},\qquad
\sin2\vartheta=\frac{175321765}{190096813},
\end{equation}
so that $\vartheta\approx0.18684\pi$. The measurements are coplanar and projective,
\begin{equation}
A_x=c^A_x\sigma_z+s^A_x\sigma_x,\qquad
B_y=c^B_y\sigma_z+s^B_y\sigma_x,
\end{equation}
with exact rational direction cosines
\begin{align}
c^A_0&=\frac{51113559740}{51657335501},
&
s^A_0&=-\frac{7475581701}{51657335501},\\
c^A_1&=\frac{96445239477}{1181952062525},
&
s^A_1&=\frac{1178010608564}{1181952062525},\\
c^B_0&=\frac{409309620877}{455458056085},
&
s^B_0&=\frac{199769054436}{455458056085},\\
c^B_1&=\frac{99679964499}{186720400949},
&
s^B_1&=-\frac{157887342140}{186720400949}.
\end{align}
Bob's overlap is the exact rational $c_0=c^B_0c^B_1+s^B_0s^B_1\approx0.1088735062$, recorded in the ancillary archive~\cite{anc}.

Writing $C=\cos2\vartheta$ and $s=\sin2\vartheta$,
the behaviour is
\[
\alpha_x=C c^A_x,\qquad
\beta_y=C c^B_y,\qquad
E_{xy}=c^A_x c^B_y+s\,s^A_x s^B_y.
\]

The eight coordinates of $\vec q$ are the exact rationals recorded in the ancillary archive~\cite{anc}. To ten decimal places,
\begin{equation}\label{eq:qex}
\begin{aligned}
\alpha_0&=0.3824626699,&
\alpha_1&=0.0315403028,\\
\beta_0&=0.3473669462,&
\beta_1&=0.2063483637,\\
E_{00}&=0.8306767176,&
E_{01}&=0.6410838222,\\
E_{10}&=0.4765022889,&
E_{11}&=-0.7336982955.
\end{aligned}
\end{equation}

All four marginals are nonzero, with
$|\alpha_0|\neq|\alpha_1|$ and
$|\beta_0|\neq|\beta_1|$.
The behaviour is also not invariant under
exchange of the two parties.
Its canonical CHSH value is approximately
$2.6819611242$, so it is nonlocal.
Extremality will follow below from the
supporting expressions and their uniqueness.

\subsubsection{The supporting hexagon}

All four fixed-overlap constraints are saturated
at $\vec q$.
The four tangent expressions, measurement
stationarity, and sixteen local comparisons
therefore define the candidate cone
$\mathcal C(\vec q)$.
Imposing $\vec g\cdot\vec q=1$ gives a
two-dimensional candidate section.

Its vertices are enumerated in exact rational
arithmetic by intersecting pairs of boundary
constraints in this affine plane and retaining
the feasible solutions.
There are six vertices,
$\vec V_0,\ldots,\vec V_5$, listed in
Table~\ref{tab:example}.
Each has strictly positive tangent weights
and saturates exactly two local deterministic
comparisons.
The exact coefficients, tangent weights,
and stationarity data are stored in the
archive~\cite{anc}.

\begin{table*}[t]
\caption{The six vertices of the normalized
supporting family at the asymmetric behaviour
of Eq.~\eqref{eq:qex}, together with two interior
expressions.
The coefficients are rounded to eight decimal
places; the defining exact rational values are
given in the archive~\cite{anc}.
The expression $\vec g_\star$ is the vertex
barycenter, and
$\vec g_{\star\star}=(\vec V_0+\vec g_\star)/2$.
Each vertex is also maximized by exactly two
local deterministic behaviours.
The two interior expressions have strict
quantum advantages and uniquely maximize at
$\vec q$.}
\label{tab:example}
\begin{center}
\footnotesize
\setlength{\tabcolsep}{4pt}
\begin{tabular}{lrrrrrrrr}
\toprule
 & $g_{\alpha_0}$ & $g_{\alpha_1}$ & $g_{\beta_0}$ & $g_{\beta_1}$ & $g_{E_{00}}$ & $g_{E_{01}}$ & $g_{E_{10}}$ & $g_{E_{11}}$ \\
\midrule
$\vec V_0$ & $0.09412922$ & $0.45836579$ & $-0.12343668$ & $0.40898770$ & $0.30280066$ & $0.23883354$ & $0.18576544$ & $-0.56544568$ \\
$\vec V_1$ & $-0.10659510$ & $0.41229536$ & $0.06127760$ & $0.49478278$ & $0.28929315$ & $0.21234846$ & $0.21592407$ & $-0.57932632$ \\
$\vec V_2$ & $0.17844917$ & $-0.04133620$ & $0.05964728$ & $-0.06690132$ & $0.38217090$ & $0.30624152$ & $0.28067640$ & $-0.37973265$ \\
$\vec V_3$ & $0.33352053$ & $0.28208518$ & $-0.22760819$ & $0.11072401$ & $0.35173994$ & $0.29549292$ & $0.19772127$ & $-0.46875821$ \\
$\vec V_4$ & $-0.19675518$ & $0.18949545$ & $0.24574765$ & $0.36205440$ & $0.31192120$ & $0.22236683$ & $0.27135931$ & $-0.51551985$ \\
$\vec V_5$ & $0.31967096$ & $0.30829185$ & $-0.22962115$ & $0.14154496$ & $0.34663577$ & $0.29047919$ & $0.19371169$ & $-0.47989966$ \\
\midrule
$\vec g_\star$ & $0.10373660$ & $0.26819957$ & $-0.03566558$ & $0.24186542$ & $0.33076027$ & $0.26096041$ & $0.22419303$ & $-0.49811373$ \\
$\vec g_{\star\star}$ & $0.09893291$ & $0.36328268$ & $-0.07955113$ & $0.32542656$ & $0.31678047$ & $0.24989698$ & $0.20497923$ & $-0.53177970$ \\
\bottomrule
\end{tabular}
\end{center}
\end{table*}

The independent quantum-bound certificates
of Appendix~\ref{app:lossy} establish
\[
h_{\mathcal Q}(\vec V_i)
=
\vec V_i\cdot\vec q
=
1
\qquad (i=0,\ldots,5).
\]
Every convex combination of the vertices
therefore also supports $\mathcal Q$ at
$\vec q$.

Define the two interior expressions
\[
\vec g_\star
=
\frac16\sum_{i=0}^{5}\vec V_i,
\qquad
\vec g_{\star\star}
=
\frac12(\vec V_0+\vec g_\star).
\]
Their local bounds are strictly below one.
The uniqueness results of
Refs.~\cite{SunWangYu2026,LiuZingarofaloBB2026}
then imply that $\vec q$ is their unique
quantum maximizer.
Consequently, $\vec q$ is a nonlocal extreme
behaviour.

Theorem~\ref{thm:dualcone} now gives
\[
N_{\mathcal Q}(\vec q)=\mathcal C(\vec q).
\]
Thus the hexagon is the complete normalized
supporting family, not merely a collection
of selected Bell expressions.
Its vertices do not have unique maximizers,
because each also attains its quantum bound
at local deterministic behaviours.

\subsubsection{Detection thresholds}

Let
\[
\mathcal P(\vec q)
=
\{\vec g\in N_{\mathcal Q}(\vec q):
\vec g\cdot\vec q=1\}
=
\operatorname{conv}\{\vec V_0,\ldots,\vec V_5\}.
\]
For a recorded behaviour $\vec p$, define
the largest local violation obtainable within
this family by
\[
M(\vec p)
=
\max_{\vec g\in\mathcal P(\vec q)}
\bigl[
\vec g\cdot\vec p-h_{\mathrm L}(\vec g)
\bigr].
\]
A positive value means that some expression
maximized by the prescribed ideal behaviour
still detects nonlocality of $\vec p$.
A nonpositive value means that none does.

Although $h_{\mathrm L}(\vec g)$ is not linear
in $\vec g$, this optimization is a linear
program.
Write
$\vec g=\sum_i\lambda_i\vec V_i$ and introduce
a variable $\nu$ for the violation margin.
Then
\begin{equation}
\label{eq:example-loss-lp}
\begin{aligned}
M(\vec p)=\max_{\lambda_i,\nu}\quad
&\nu\\
\text{subject to}\quad
&\lambda_i\geq0,\qquad \sum_i\lambda_i=1,\\
&\nu\leq
\sum_i\lambda_i\vec V_i\cdot(\vec p-\vec d)\\
&\qquad\text{for every local deterministic }\vec d.
\end{aligned}
\end{equation}
There are only sixteen inequalities of the
last type.

Here we vary the detector efficiencies while
keeping the ideal behaviour $\vec q$ fixed.
A missed detection is recorded as $+1$,
and every trial is retained.
For efficiencies $\eta_A,\eta_B$, the recorded
behaviour $\vec p=L_{\eta_A,\eta_B}(\vec q)$
has
\[
\alpha'_x=\eta_A\alpha_x+1-\eta_A,
\qquad
\beta'_y=\eta_B\beta_y+1-\eta_B,
\]
and
\[
\begin{aligned}
E'_{xy}
={}&\eta_A\eta_B E_{xy}
+\eta_A(1-\eta_B)\alpha_x\\
&+(1-\eta_A)\eta_B\beta_y
+(1-\eta_A)(1-\eta_B).
\end{aligned}
\]

For Alice-only loss, $\eta_A=\eta$ and
$\eta_B=1$, the optimal critical efficiency
within $\mathcal P(\vec q)$ is approximately
$0.757617$.
The vertex barycenter $\vec g_\star$ has
critical efficiency approximately $0.792657$.
The certified reduction is at least
$0.035040$.
At the exact critical efficiency the optimal
violation margin is zero; the corresponding
test violates its local bound above that
efficiency.

The optimal test is the stored point of the
hexagon whose local bound is attained at the
three local deterministic assignments
\[
(1,1,1,1),\qquad
(1,1,1,-1),\qquad
(1,-1,1,1).
\]
Its exact coefficients and the certified
threshold data are given in the
archive~\cite{anc}.

Optimality within the family follows from a
local behaviour $\vec\ell$ satisfying the
conditions below.
The linear program also provides a way to
certify that no other expression in the
supporting family has a lower critical
efficiency.
For any local behaviour $\vec\ell$ and any
$\vec g\in\mathcal P(\vec q)$,
\[
\vec g\cdot\vec p-h_{\mathrm L}(\vec g)
\leq
\vec g\cdot(\vec p-\vec\ell)
\leq
\max_i\vec V_i\cdot(\vec p-\vec\ell).
\]
Thus a local behaviour satisfying
\[
\vec V_i\cdot(\vec p-\vec\ell)\leq0
\qquad\text{for every }i
\]
certifies $M(\vec p)\leq0$.

At the certified family threshold, the stored
dual certificate gives a local behaviour
$\vec\ell$, expressed as a convex combination
of local deterministic behaviours, such that
\[
\vec V_i\cdot(\vec p-\vec\ell)=0
\qquad\text{for all six vertices}.
\]
Together with the test attaining zero
violation margin, this proves optimality
within the complete supporting family.
For Alice-only loss, the recorded behaviour
is affine in $\eta$, and every normalized
supporting expression has value one at
$\eta=1$.
Once such an expression violates its local
bound, its violation therefore persists as
$\eta$ increases.
The certificate consequently identifies the
critical efficiency: no expression in the family
violates at or below this value, while the
selected expression violates above it.

The same stored expression is optimal for
the Bob-only, symmetric, and calibrated
unequal-efficiency cases examined in the
archive~\cite{anc}.
This statement concerns those specified loss
trajectories, not arbitrary detector
imperfections.

For comparison, ordinary CHSH has a lower
critical efficiency along the same loss trajectory.
The optimization above preserves the ideal
maximizer: every permitted expression must
belong to $N_{\mathcal Q}(\vec q)$.
It is therefore different from asking whether
the recorded behaviour is nonlocal at all.

Along the same Alice-only loss trajectory,
ordinary CHSH remains violated down to a
critical efficiency of approximately
$0.656828$.
CHSH does not belong to the supporting family
at $\vec q$.
At this lower threshold the recorded behaviour
enters the local set, as verified using the
positivity constraints and the eight CHSH
inequalities that characterize locality in
this scenario.

Thus there is a range of efficiencies in
which the recorded behaviour is nonlocal
but no Bell expression having $\vec q$ as
an ideal maximizer detects it.
The hexagon construction makes this
restriction explicit and allows optimality
within the prescribed family to be certified.

\end{document}